\documentclass{article}
\usepackage{epsfig}
\usepackage{graphicx}
\usepackage[centertags]{amsmath}
\usepackage{amsfonts}
\usepackage{amssymb}
\usepackage{amsthm}
  
\begin{document}

\title{Affinely-Rigid Body and Oscillatory Dynamical Models on ${\rm GL}(2,\mathbb{R})$}

\author{A. Martens, J. J. S\l awianowski\\
Institute of Fundamental Technological Research,\\
Polish Academy of Sciences,\\
 Pawi\'{n}skiego 5B, 02-106 Warszawa, Poland\\
e-mails: amartens@ippt.gov.pl, jslawian@ippt.gov.pl}

\maketitle
\begin{abstract}     
Discussed is a model of the two-dimensional affinely-rigid body with
the double dynamical isotropy. We investigate the
systems with potential energies for which the variables can
be separated. The special stress is laid on the model of the
harmonic oscillator potential and certain anharmonic alternatives.
Some explicit solutions are found on the
classical, quasiclassical (Bohr-Sommerfeld) and quantum
levels.
\end{abstract}

\section{Introduction}

The mechanics of an affinely-rigid body was discussed in various
aspects in \cite{2}, \cite {7}--\cite{16}. In this paper we intend to
investigate qualitatively the doubly-isotropic dynamical models in
two dimensions, having in view applications in macroscopic
elasticity and the theory of molecular vibrations. We expect also
applications in dynamics of nanotubes; more precisely, we mean vibrations
of their transversal cross-sections. On the
classical level our models are completely integrable and may show some
degeneracy properties following from hidden symmetries. In the
two-dimensional theory there exists a relatively wide class of isotropic
potentials which admit analytical calculations based on the
separation of variables method \cite{7,8}. In this paper the special
stress is laid on the model of the harmonic oscillator potential 
and certain anharmonic models.
The action-angle analysis and discussion of degeneracy as well as the
quasiclassical Bohr-Sommerfeld quantization are also presented.
Next we discuss the Schr\"{o}dinger quantization
procedure for such an object. We follow the standard procedure of
quantization in Riemannian manifolds \cite{4}, i.e. we use the
$L^{2}$-Hilbert space of wave functions in the sense of the usual
Riemannian measure (volume element). Some explicit solutions are
found using the Sommerfeld polynomial method \cite{6,6a}.

Certain of our ideas are somehow related to those underlying the papers
\cite{3b,3c}.

\section{Geometric description of the affinely-rigid body}

We are given two Euclidean spaces $(N,U,\eta)$ and $(M,V,g)$,
respectively the material and physical spaces. Here $N$ and $M$ are
the basic point spaces, $U$ and $V$ are their linear translation
spaces, and $\eta\in U^{\ast}\otimes U^{\ast}$, $g\in
V^{\ast}\otimes V^{\ast}$ are their metric tensors. The space $N$ is
used for labelling the material points, and elements of $M$ are
geometric spatial points.

The configuration space of the affinely-rigid body 
\[
Q:={\rm AfI}(N,M)
\]
consists of affine isomorphisms of $N$ onto $M$. The material
labels $a\in N$ are parametrized by Cartesian coordinates
$a^{K}$ (Lagrange variables). Cartesian coordinates in $M$ will be
denoted by $y^{i}$ and the corresponding geometric points by $y$.
The configuration $\Phi\in Q$ is to be understood in such a way
that the material point $a\in N$ occupies the spatial position
$y=\Phi(a)$.

Let $\overline{\mu}$ denote the co-moving (Lagrangian) mass distribution in
$N$; obviously, it is constant in time. Lagrange coordinates
$a^{K}$ in $N$ will be always chosen in such a way that their
origin $a^{K}=0$ coincides with the centre of mass $\mathcal{C}$:
\[
\int a^{K} d \overline{\mu}(a)=0. 
\]
The configuration space may be
identified then with $ M\times {\rm LI}(U,V)$, 
\[
Q={\rm AfI}(N,M)\simeq
 M\times {\rm LI}(U,V)= M\times Q_{\rm int},
\]
where ${\rm LI}(U,V)$ denotes
the manifold of all linear isomorphisms of $U$ onto $V$. The
Cartesian product factors refer respectively to the translational
motion $(M)$ and the internal or relative motion $({\rm LI}(U,V))$. The motion is
described as a continuum of instantaneous configurations:
\begin{equation}\label{EQa'}
\Phi (t,a)^{i}=\phi^{i}{}_{K}(t)a^{K}+x^{i}(t),
\end{equation}
where $x(t)$ is the centre of mass position and $\phi(t)$
tells us how constituents of the body are placed with respect to
the centre of mass. The quantities $\left(x^{i},\phi^{i}{}_{K}\right)$ are
our generalized coordinates. 

Obviously, if we put $U=V=\mathbb{R}^{n}$, then $Q_{{\rm int}}$ reduces to 
${\rm GL}(n,\mathbb{R})$ and $Q$ becomes the semi-direct product 
$\mathbb{R}^{n}\times_{s}{\rm GL}(n,\mathbb{R})$; $\mathbb{R}^{n}$ is then
interpreted as an Abelian group with addition of vectors as a group operation.

Inertia of affinely-constrained systems of material points is
described by two constant quantities:
\[
m=\int  d \overline{\mu}(a),
\qquad J^{KL}=\int a^{K}a^{L} d \overline{\mu}(a), 
\]
i.e. the total mass
$m$ and the co-moving second-order moment $J\in U\otimes U$. More
precisely, it is so in the usual theory based on the d'Alembert
principle, when the kinetic energy is obtained by summation
(integration) of usual (based on the metric $g$) kinetic energies
of constituents \cite{8}--\cite{13},
\[
T=\frac{1}{2}g_{ij}\int \frac{\partial \Phi^{i}}{\partial
t}\frac{\partial \Phi^{j}}{\partial t} d \overline{\mu}(a).
\]
Substituting
to this general formula the above affine constraints (\ref{EQa'}) we
obtain:
\[
T=T_{\rm tr}+T_{\rm int}=\frac{m}{2}g_{ij}\frac{
d x^{i}}{ dt} \frac{d x^{j}}{d t}+\frac{1}{2}g_{ij}\frac{
d\phi^{i}_{A}}{d t} \frac{d \phi^{i}_{B}}{d t}J^{AB}.
\]
Certainly, if we analytically identify $U$ and $V$ with
$\mathbb{R}^{n}$ and ${\rm LI}(U,V)$ with ${\rm GL}(n,\mathbb{R})$, then 
\[
T_{\rm
int}=\frac{1}{2}{\rm Tr}\left(\dot{\phi}^{T}\dot{\phi}J\right).
\] 

\section{Some two-dimensional problems}

Now, let us discuss the two-dimensional affinely-rigid body. Considered is
a discrete or continuous system of material points subject to
constraints according to which during any admissible motion all
affine relations between constituents of the body are invariant
(the material straight lines remain straight lines, their
parallelism is conserved, and all mutual ratios of segments placed
on the same straight lines are constant). The conception of the
affinely-rigid body is a generalization of the usual
metrically-rigid body, in which during any admissible motion all
distances (metric relations) between its constituents are constant
\cite{1}. We do not take into account the motion of the centre of
mass. When translational motion is neglected, the configuration
space $Q$ may be analytically identified with the linear group
${\rm GL}(2,\mathbb{R})$, i.e., the group of non-singular real $2\times 2$
matrices. The most adequate description of degrees 
of freedom is that based on the two-polar decomposition of matrices:
\begin{equation}\label{EQ2.0}
\phi =ODR^{T},
\end{equation}
 where
\[
O= \left[\begin{array}{cc}
\cos \varphi & -\sin \varphi \\
\sin \varphi & \cos \varphi
\end{array}\right]
,\ D=\left[\begin{array}{cc}
D_{1} & 0 \\
0 & D_{2}
\end{array}\right],
\  R=\left[\begin{array}{cc}
\cos \psi & -\sin \psi \\
\sin \psi & \cos \psi
\end{array}\right].
\]
This decomposition is connected with the algebraic Gram-Schmid
orthogonalization. It is also know in literature as the 
"singular value decomposition".
The matrices $O, R\in {\rm SO}(2,\mathbb{R})$ are orthogonal
($O^{T}O=R^{T}R={\rm Id}$, ${\rm det} O={\rm det} R=1$), $D$ is diagonal and positive.
The orthogonal group ${\rm SO}(2,\mathbb{R})$ is a commutative
group of plane rotations. Spatial rotations are described by the
action of ${\rm SO}(2,\mathbb{R})$ on ${\rm GL}(2,\mathbb{R})$ through the
left regular translations, material rotations are represented by
the action of the rotation subgroup through the right
multiplication. In the non-degenerate case $(D_{1}\neq D_{2})$, 
the decomposition (\ref{EQ2.0}) is unique up to
the permutation of the diagonal elements of $D$ accompanied by the
simultaneous multiplying of $O$ and $R$ on the right-side by the
appropriate special orthogonal matrices
(ones having in each row and column zeros but once $\pm 1$ as elements). 
This implies that the
potential energy of doubly isotropic models depends only on $D$
and is invariant with respect to the permutations of its
nonvanishing matrix elements \cite{8}. The deformation invariants
$D_{1}$, $D_{2}$ are important mechanical quantities. They are
scalar measures of deformation, i.e. tell us how strongly the body
is deformed, but do not contain any information concerning the
orientation of deformation in the physical or material space. The
orthogonal matrices $O$ and $R$ describe the space and body
orientations of the strain. Incidentally, let us mention that the 
complexification of ${\rm GL}(2,\mathbb{R})$ to ${\rm GL}(2,\mathbb{C})$ 
and then the restriction to the other, completely opposite (because 
compact), real form ${\rm U}(2)$ sheds some light on our model and establishes
also certain kinship with the three-dimensional rigid body.

We shall consider only highly symmetric model, where $J$ is
isotropic, i.e., its matrix has the form $\mu I$, $\mu$ denoting a
positive constant, and $I$ is the $2\times 2$ identity matrix. The
isotropic kinetic energy is as follows:
\begin{eqnarray}\label{EQ2.0a}
T&=&\frac{\mu }{2}\left[ \left( D_{1}{}^{2}+ D_{2}{}^{2}\right)\left(
\left( \frac{ d\varphi }{dt}\right) ^{2}+ \left( \frac{ d\psi
}{dt}\right) ^{2}\right)-4D_{1}D_{2}\frac{d\varphi }{dt}\frac{d\psi
}{dt}\right.\nonumber\\
&+&\left. \left( \frac{dD_{1}}{dt}\right) ^{2}+\left(
\frac{dD_{2}}{dt}\right)^{2} \right].
\end{eqnarray}
The matrices $O$ and $R$ do not enter into this equation, hence
the angles $\varphi $, $\psi $ are cyclic variables. In these
coordinates the Hamilton-Jacobi equation is non-separable even in
the interaction-free case. However, the separability becomes
possible in new variables, obtained by the $\pi /
4$-rotation in the plane of the deformation invariants $D_{1}$,
$D_{2}$  and by an appropriate modification of the angular variables.
Thus, we introduce the following new coordinates:
\[
\alpha =\frac{1}{\sqrt{2}}\left( D_{1}+D_{2}\right), \quad \beta
=\frac{1}{\sqrt{2}}\left( D_{1}-D_{2}\right), \quad \eta = \varphi
-\psi, \quad \gamma =\varphi +\psi.
\]
In the macroscopic, phenomenological elasticity theory
 $D_{1}>0$ , $\
 D_{2}>0$, thus, $\alpha >0$, $\ \left| \beta \right| <\alpha $. However,
describing discrete or finite systems of material points (e.g.
molecules), one can admit singular and mirror-reflected
configurations. Then, to some extent $D_{1}$, $ D_{2}$, $\alpha
$, $\beta $ may be arbitrary. The kinetic energy becomes then
\begin{equation}\label{EQ2.0b}
T=\frac{\mu }{2}\left[ \alpha ^{2}\left( \frac{d\eta }{dt}\right)
^{2}+\beta ^{2}\left( \frac{d\gamma }{dt}\right) ^{2}+\left(
\frac{d\alpha }{dt}\right) ^{2}+\left( \frac{d\beta }{dt}\right)
^{2}\right].
\end{equation}
This form is both diagonal and separable. The classical St\"{a}ckel 
theorem leads to the following general form of
separable potentials:
\begin{equation}\label{EQ2.1}
V(\varphi ,\psi ,\alpha ,\beta )=\frac{V_{\eta }\left( \varphi -\psi
\right)
 }{\alpha ^{2}}+\frac{V_{\gamma }\left( \varphi +\psi \right) }
 {\beta ^{2}} +V_{\alpha
}(\alpha )+V_{\beta }(\beta ).
\end{equation}
In this formula $V_{\eta }$, $V_{\gamma }$, $V_{\alpha }$,
$V_{\beta }$ are arbitrary (but regular enough) functions of a
single variable (indicated as an argument). We consider
doubly-isotropic models in which the potential energy does not
depend on variables $\varphi $, $\psi $ (equivalently $\eta $,
$\gamma $), i.e. $V_{\eta }=0$ and $V_{\gamma }=0$. Performing the
Legendre transformation we obtain the corresponding Hamiltonian 
$H=H_{\alpha}+H_{\beta}$ in the form:
\begin{equation}\label{EQ2.1a}
H=\frac{1}{2\mu }\left( \frac{(p_{\varphi
}-p_{\psi })^{2}} {4\alpha ^{2}} +p_{\alpha }{}^{2}\right)
+\frac{1}{2\mu }\left( \frac{(p_{\varphi }+ p_{\psi })^{2}}{4\beta
^{2}}+p_{\beta }{}^{2}\right)+V_{\alpha }(\alpha )+V_{\beta }(\beta
),
\end{equation}
where $p_{\varphi }$, $p_{\psi }$, $p_{\alpha }$, $p_{\beta }$ are
the canonical momenta conjugate to $\varphi $, $\psi $, $ \alpha $,
$\beta $, respectively, and
\begin{eqnarray}\label{EQ2.1b}
H_{\alpha}&=&\frac{1}{2\mu}\left(\frac{(p_{\varphi}-p_{\psi})^{2}}{4\alpha^{2}}+
p_{\alpha}{}^{2}\right)+V_{\alpha }(\alpha ), \nonumber \\
H_{\beta}&=&\frac{1}{2\mu}\left(\frac{(p_{\varphi}+p_{\psi})^{2}}{4\beta^{2}}+
p_{\beta}{}^{2}\right)+V_{\beta }(\beta ).
\end{eqnarray}
The quantities $H_{\alpha}, H_{\beta}, p_{\varphi}, p_{\psi}$ form a
Poisson-involutive system of constants of motion.

The stationary Hamilton-Jacobi equation has the following form:
\begin{equation}
\left(\frac{1}{4\alpha^{2}}+\frac{1}{4\beta^{2}}\right)\left(\left(\frac{\partial
S }{\partial \varphi}\right)^{2}+\left(\frac{\partial S }{\partial
\psi}\right)^{2}\right)+\left(\frac{1}{2\beta^{2}}-\frac{1}{2\alpha^{2}}\right)
\frac{\partial^{2} S }{\partial \varphi\partial \psi}
\end{equation}
\[
+\left(\frac{\partial S }{\partial
\alpha}\right)^{2}+\left(\frac{\partial S }{\partial
\beta}\right)^{2}=2\mu\left(E-(V_{\alpha }(\alpha )+V_{\beta
}(\beta ))\right),
\]
where $E$ is a fixed value of the energy. Due to the fact that the
variables $\varphi $, $\psi $\ have the cyclic character, we may
write:
\[
S=S_{\varphi }(\varphi)+ S_{\psi}(\psi) +S_{\alpha}(\alpha)+S_{\beta
}(\beta)=a\varphi +b\psi +S_{\alpha}(\alpha)+S_{\beta}(\beta)
\]
and the action variables are as follows:
\begin{equation}
J_{\varphi }=\oint p_{\varphi}d\varphi =2\pi a,\quad J_{\alpha}= \pm
\oint \sqrt{2\mu \left( E_{\alpha }-V_{\alpha}(\alpha)\right)
-\frac{\left(J_{\varphi}- J_{\psi}{}\right) ^{2}}{16\pi^{2}\alpha
^{2}}}d\alpha,
\end{equation}
\begin{equation}\label{EQ2.2}
J_{\psi }=\oint p_{\psi }d\psi =2\pi b,\quad J_{\beta }= \pm \oint
\sqrt{2\mu \left( E_{\beta }-V_{\beta}(\beta)\right) -\frac{\left(
J_{\varphi}+ J_{\psi}{}\right) ^{2}} {16\pi^{2}\beta ^{2}}}\
d\beta,
\end{equation}
where $E_{\alpha }$, $E_{\beta }$, $a$, $b$\ are separation
constants. 

\noindent{\bf Remark.} Let us observe that the isotropic kinetic energy
\begin{equation}\label{EQ2.2a}
T=\frac{\mu}{2}{\rm Tr}\left(\dot{\phi}^{T}\dot{\phi}\right)
\end{equation}
may be simply written as
\begin{equation}\label{EQ2.2b}
T=\frac{\mu}{2}\left(\dot{x}^{2}+\dot{y}^{2}+\dot{z}^{2}+\dot{u}^{2}\right),
\end{equation}
where $x, y, z, u$ are simply the matrix elements of $\phi$,
\begin{equation}\label{EQ2.2c'}
\phi=\left[\begin{array}{cc}
x & y \\
z & u \\
\end{array}\right].
\end{equation}
This is formally the expression for the material point with the mass
$\mu$ in $\mathbb{R}^{4}$ or the quadruple of such material points in 
$\mathbb{R}$. However, in the mechanics of deformable bodies these generalized
coordinates are not very useful for dynamical models.

It is both convenient and instructive to use also other generalized
coordinates in the affine kinematics. We mean coordinates in which the
problem is separable; as mentioned, the separability in various coordinates 
corresponds geometrically to some degeneracy of the problem. And besides,
those coordinates suggest some modifications of the potential $V$
leading to new models of deformative dynamics, more realistic than the 
harmonic oscillator and at the same time admitting also some analytical
treatment. As expected, in doubly isotropic models the most natural 
candidates are to be sought among orthogonal coordinates on the plane
of the deformation invariants $(D_{1}$, $D_{2})$. The most natural of them 
are just the variables $\alpha$, $\beta$ introduced above: they are obtained from
$D_{1}$, $D_{2}$ by the rotation by $\pi /4$ in $\mathbb{R}^{2}$. Together with
the modified angular variables $\eta$, $\gamma$ they provide a system of 
$T$-orthogonal coordinates in $\mathbb{R}^{4}$, i.e., in the space
of variables $x$, $y$, $z$, $u$. To be more precise, they are orthogonal
coordinates for the metric element $dx^{2}+dy^{2}+dz^{2}+du^{2}$
on which the kinetic energy $T$ is based. And moreover, as said above, they are
the nice separation variables for $T$ in the St\"{a}ckel sense. Other natural 
$T$-separating variables are obtained as some byproducts of $\alpha$, $\beta$.
The most natural of them are polar variables in the $\mathbb{R}^{2}$-plane
of the pairs $(\alpha, \beta)$. In certain problems it is analytically
convenient to use the modified "polar" variables $r$, $\vartheta$ given by
\[
\alpha=\sqrt{r}\cos\frac{\vartheta}{2}, \quad \beta=\sqrt{r}\sin\frac{\vartheta}{2}.
\]
Obviously, the "literal" polar variables $\rho$, $\epsilon$ are defined by
\[
\alpha=\rho\cos\epsilon, \quad \beta=\rho\sin\epsilon; \quad \rho=\sqrt{r}, 
\quad \epsilon=\frac{\vartheta}{2}.
\]

The natural metric on the manifold of $2 \times 2$ matrices,
\[
ds^{2}={\rm Tr}\left(d\phi^{T}d\phi\right)=dx^{2}+dy^{2}+dz^{2}+du^{2}, 
\]
becomes then
\begin{eqnarray*}
ds^{2}&=&r\cos^{2}\frac{\vartheta}{2}d\eta^{2}+r\sin^{2}\frac{\vartheta}{2}d\gamma^{2}+
\frac{1}{4r}dr^{2}+\frac{r}{4}d\vartheta^{2}\\
&=& d\rho^{2}+\rho^{2}d\epsilon^{2}+\rho^{2}\cos^{2}\epsilon \ d\eta^{2}+
\rho^{2}\sin^{2}\epsilon \ d\gamma^{2}\\
&=&d\rho^{2}+\frac{1}{4}\rho^{2}d\vartheta^{2}+\rho^{2}\cos^{2}\frac{\vartheta }{2}\ 
d\eta^{2}+\rho^{2}\sin^{2}\frac{\vartheta}{2} \ d\gamma^{2}.
\end{eqnarray*}
Evidently, kinetic energy is then expressed as follows
\begin{eqnarray*}
T&=&\frac{\mu}{2}\left(\frac{1}{4r}\left(\frac{dr}{dt} \right)^{2}+
\frac{r}{4}\left(\frac{d\vartheta}{dt} \right)^{2}+
r\cos^{2}\frac{\vartheta}{2}\left(\frac{d\eta}{dt} \right)^{2} +
r\sin^{2}\frac{\vartheta}{2}\left(\frac{d\gamma}{dt} \right)^{2} \right)\\
&=& \frac{\mu}{2}\left(\left(\frac{d\rho}{dt} \right)^{2}+
\rho^{2}\left(\frac{d\epsilon}{dt} \right)^{2}+
\rho^{2}\cos^{2}\epsilon\left(\frac{d\eta}{dt}\right)^{2} +
\rho^{2}\sin^{2}\epsilon\left(\frac{d\gamma}{dt} \right)^{2} \right)\\
&=& \frac{\mu}{2}\left(\left(\frac{d\rho}{dt}\right)^{2}+
\frac{1}{4}\rho^{2}\left(\frac{d\vartheta}{dt}\right)^{2}+
\rho^{2}\cos^{2}\frac{\vartheta}{2}\left(\frac{d\eta}{dt} \right)^{2} +
\rho^{2}\sin^{2}\frac{\vartheta}{2}\left(\frac{d\gamma}{dt} \right)^{2} \right).
\end{eqnarray*}
The above crowd of expressions is due to the fact that different conventions
are better suited to different analogies: the two-dimensional homogeneously
deformable body and three-dimensional spherical top with dilatations.
Physically we are interested here in the first problem, however, certain
aspects of the second one (spherical top with dilatations) are formally useful
and the mysterious link between them is interesting in itself.

Let us notice that $(r, \vartheta)$ may be interpreted as polar coordinates 
in the two-dimensional space of quantities $2D_{1}D_{2}$, $D_{1}{}^{2}-D_{2}{}^{2}$,
\begin{equation}\label{EQ2.2y}
2D_{1}D_{2}=r\cos\vartheta, \quad D_{1}{}^{2}-D_{2}{}^{2}=r\sin\vartheta,
\end{equation}
or, inverting these formulas,
\begin{equation}\label{EQ2.2x}
r=\rho^{2}=D_{1}{}^{2}+D_{2}{}^{2}, \quad  \tan\vartheta=\tan(2\epsilon)=\frac{1}{2}
\left(\frac{D_{1}}{D_{2}}-\frac{D_{2}}{D_{1}}\right).
\end{equation}
Therefore, $\vartheta$ refers to the shear degrees of freedom, whereas $r=\rho^{2}$
is some kind of the measure of size. More precisely, dilatation is measured by the product
$D_{1}D_{2}$, thus,
\begin{equation}\label{EQ2.2z}
r=\frac{2D_{1}D_{2}}{\cos\vartheta}
\end{equation}
contains an "admixture" of the shear parameter $\vartheta$. Nevertheless, just like 
$D_{1}D_{2}$ it is a homogeneous function of degree $2$ of $(D_{1}, D_{2})$.
The shear parameter $\vartheta$ is evidently a homogeneous function of degree zero.

It is also convenient to parametrize deformation invariants as follows:
\[
D_{1}={\rm exp}\left(\frac{a+b}{2}\right), \quad D_{2}={\rm exp}\left(\frac{a-b}{2}\right).
\]
Then
\[
\alpha=\frac{1}{\sqrt{2}}(D_{1}+D_{2})=\sqrt{2}e^{\frac{a}{2}}\cosh\frac{b}{2}, \quad \beta=
\frac{1}{\sqrt{2}}(D_{1}-D_{2})=\sqrt{2}e^{\frac{a}{2}}\sinh\frac{b}{2},
\]
\[
D_{1}D_{2}=e^{a}, \quad D_{1}{}^{2}+D_{2}{}^{2}=2e^{a}\cosh b, \quad D_{1}{}^{2}-
D_{2}{}^{2}=2e^{a}\sinh b, \quad \frac{D_{1}}{D_{2}}=e^{b},
\]
\[
\sin\vartheta=\tanh b, \quad  \cos\vartheta=
\frac{1}{\cosh b}, \quad \tan\vartheta=\sinh b.
\]
These simple formulas shed some light onto the link between two-dimensional
homogeneously deformable body and three-dimensional top. Nevertheless, this 
link is still rather mysterious and obscure.

For the completeness let us also mention about other orthogonal 
coordinates on the plane of deformation invariants:
\begin{itemize}
\item[$(i)$] Elliptic variables $(\kappa, \lambda)$, where
\[
\alpha=\sqrt{2}\cosh\kappa\cos\lambda, \quad \beta=\sqrt{2}\sinh\kappa\sin\lambda.
\]
\item[$(ii)$] Parabolic variables $(\xi, \delta)$, where
\[
\alpha=\frac{1}{2}\left(\xi^{2}-\delta^{2}\right), \quad \beta=\xi\delta.
\]
\item[$(iii)$] Two-polar variables $(e, f)$, where
\[
\alpha=\frac{c\sinh e}{\cosh e-\cos f}, \quad \beta=\frac{c \sin f}{\cosh e-\cos f},
\]
\end{itemize}
and $c$ is a constant.

For our analysis of the deformative motion the parabolic $(\xi, \delta)$ and two-polar
variables $(e, f)$ are non-useful, because the corresponding Hamilton-Jacobi equations are non-separable
even in the non-physical geodetic models, i.e., ones with vanishing potentials.
In the elliptic coordinates $(\kappa, \lambda)$ the metric underlying the kinetic 
energy takes on the form:
\begin{eqnarray*}
ds^{2}&=&{\rm Tr}\left(d\phi^{T}d\phi\right)=\left(\cosh^{2}\kappa-\cos^{2}\lambda
\right)d\kappa^{2}\\
&+&\left(\cosh^{2}\kappa-\cos^{2}\lambda
\right)d\lambda^{2}
+\cosh^{2}\kappa\cos^{2}\lambda d\eta^{2}+
\sinh^{2}\kappa\sin^{2}\lambda d\gamma^{2}.
\end{eqnarray*}

The general St\"{a}ckel-separable Hamiltonians $H=T+V$ in the variables 
$(\alpha, \beta, \eta, \gamma)$, $(r, \vartheta, \eta, \gamma)$ and 
$(\kappa, \lambda, \eta, \gamma)$ have respectively the form:

\begin{eqnarray}
H&=&\frac{1}{2\mu}\left(\left(p_{\alpha}{}^{2}+\frac{p_{\eta}{}^{2}}{\alpha^{2}}\right)
+\left(p_{\beta}{}^{2}+\frac{p_{\gamma}{}^{2}}{\beta^{2}}\right)\right)\nonumber\\
&+&V_{\alpha}(\alpha)+V_{\beta}(\beta)+\frac{V_{\eta}(\eta)}{\alpha^{2}}+
\frac{V_{\gamma}(\gamma)}{\beta^{2}},\\
H&=&\frac{1}{2\mu}\left(4rp_{r}{}^{2}+\frac{1}{r}\left(\frac{p_{\varphi}{}^{2}+
p_{\psi}{}^{2}+2p_{\varphi}p_{\psi}\cos\vartheta}{\sin^{2}\vartheta}
+4p_{\vartheta}{}^{2}\right)\right)\nonumber \\
&+&V_{r}(r)+\frac{V_{\vartheta}(\vartheta)}{r}+\frac{V_{\eta}(\eta)}
{r\cos^{2}\frac{\vartheta}{2}}+\frac{V_{\gamma}(\gamma)}
{r\sin^{2}\frac{\vartheta}{2}},\label{EQ2.2f}\\
H&=&\frac{1}{4\mu}\left(\frac{p_{\kappa}{}^{2}}{\left(\cosh^{2}\kappa-\cos^{2}\lambda\right)}
+\frac{p_{\lambda}{}^{2}}{\left(\cosh^{2}\kappa-\cos^{2}\lambda\right)}\right.\nonumber\\
&+&\left.\frac{p_{\eta}{}^{2}}{\cosh^{2}\kappa\cos^{2}\lambda}+\frac{p_{\gamma}{}^{2}}
{\sinh^{2}\kappa\sin^{2}\lambda}\right)\nonumber\\
&+&\frac{V_{\kappa}(\kappa)}{2\left(\cosh^{2}\kappa-\cos^{2}\lambda\right)}+
\frac{V_{\lambda}(\lambda)}{2\left(\cosh^{2}\kappa-\cos^{2}\lambda\right)}\nonumber\\
&+&\frac{V_{\eta}(\eta)}{2\cosh^{2}\kappa\cos^{2}\lambda}+\frac{V_{\gamma}(\gamma)}
{2\sinh^{2}\kappa\sin^{2}\lambda}.
\end{eqnarray}

Let us observe that, obviously,
\[
\cosh^{2}\kappa-\cos^{2}\lambda=\sinh^{2}\kappa+\sin^{2}\lambda
\]
and it does not matter what is written in the corresponding denominators above. Making
use of this fact we immediately see that when the problem is doubly isotropic, i.e.,
$V_{\eta}$, $V_{\gamma}$ are constant, then  obviously $(p_{\eta}, p_{\gamma})$, 
equivalently $(p_{\varphi}, p_{\psi})$, are constants of motion but also there 
is a separation of the Hamilton-Jacobi equation in the variables $\kappa$, $\lambda$.
Therefore, there are two additional constants of motion and the problem is integrable.
Those constants of motion are given by
\begin{eqnarray*}
K&=&\frac{h_{\kappa}\cos^{2}\lambda-h_{\lambda}\cosh^{2}\kappa}{2\left(\cosh^{2}\kappa
-\cos^{2}\lambda\right)}=\frac{h_{\kappa}\cos^{2}\lambda-h_{\lambda}\cosh^{2}\kappa}
{2\left(\sinh^{2}\kappa+\sin^{2}\lambda\right)},\\
L&=&\frac{h_{\kappa}\sin^{2}\lambda-h_{\lambda}\sinh^{2}\kappa}{2\left(\sinh^{2}\kappa
+\sin^{2}\lambda\right)}=\frac{h_{\kappa}\sin^{2}\lambda-h_{\lambda}\sinh^{2}\kappa}
{2\left(\cosh^{2}\kappa-\cos^{2}\lambda\right)},
\end{eqnarray*}
where the auxiliary quantities $h_{\kappa}$, $h_{\lambda}$ are not
constants of motion and are respectively given by 
\[
h_{\kappa}=\frac{1}{2\mu}\left(p_{\kappa}{}^{2}+2\mu V_{\kappa}-\frac{\frac{1}{4}
(p_{\varphi}-p_{\psi})^{2}+2\mu V_{\kappa}}{\cosh^{2}\kappa}+\frac{\frac{1}{4}
(p_{\varphi}+p_{\psi})^{2}+2\mu V_{\kappa}}{\sinh^{2}\kappa}\right),
\]
\[
h_{\lambda}=\frac{1}{2\mu}\left(p_{\lambda}{}^{2}+2\mu V_{\lambda}+\frac{\frac{1}{4}
(p_{\varphi}-p_{\psi})^{2}+2\mu V_{\lambda}}{\cos^{2}\lambda}+\frac{\frac{1}{4}
(p_{\varphi}+p_{\psi})^{2}+2\mu V_{\lambda}}{\sin^{2}\lambda}\right);
\]
we remember that $V_{\kappa}$, $V_{\lambda}$ are constants here.

Therefore, we have the involutive system of constants of motion (their Poisson
brackets do vanish), and 
\[
H=K+L
\]
has the vanishing Poisson brackets with all of them, i.e., with $p_{\varphi}$, $p_{\psi}$
(i.e., with $p_{\eta}$, $p_{\gamma})$, $K$, $L$. 

The elliptic coordinates and the corresponding separable models are not very interesting
for applications. From this point of view the "polar" coordinates $(r, \vartheta)$, or
equivalently $(\rho, \epsilon)$, are much more useful. The configurational metric tensor
is then expressed as follows:
\begin{eqnarray*}
ds^{2}&=&{\rm Tr}\left(d\phi^{T}d\phi\right)=\frac{1}{4r}dr^{2}+\frac{r}{4}d\vartheta^{2}
+r d\varphi^{2}-2r\cos\vartheta d\varphi d\psi
+rd\psi^{2}\\&=&d\rho^{2}+\frac{1}{4}\rho^{2}\left(d\vartheta^{2}+d(2\varphi)^{2}-
2\cos\vartheta d(2\varphi)d(2\psi)+d(2\psi)^{2}\right)\\
&=&\frac{1}{4r}\left(dr^{2}+r^{2}\left(d\Theta^{2}+d\Phi^{2}-2\cos\Theta d\Phi d\Psi
+d\Psi^{2}\right)\right),
\end{eqnarray*}
where, obviously, the doubled angles are used, $\Theta=\vartheta$, $\Phi=2\varphi$, 
$\Psi=2\psi$.
This expression is very interesting in itself. We used here three alternative systems
of symbols, each of them convenient and suggestive in some areas of applications.
It is seen that the expression
\[
d\sigma^{2}=d\Theta^{2}+d\Phi^{2}-2\cos\Theta d\Phi d\Psi +d\Psi^{2}
\]
is exactly, up to a constant multiplier, identical with the doubly-invariant (i.e.,
both left- and right-invariant) squared metric element on the rotation group
in three dimensions, ${\rm SO}(3, \mathbb{R})$, or on its covering group 
${\rm SU}(2)$. This identification is based on interpreting $\Phi, \Theta, \Psi$
as Euler angles. More precisely, to be literal in this analogy, one should change
the sign at $\Psi$, then one obtains the usual expression
\[
d\sigma^{2}=d\Theta^{2}+d\Phi^{2}+2\cos\Theta d\Phi d\Psi+d\Psi^{2}.
\]
This metric underlies the kinetic energy expression for the spherical top,
\[
T=\frac{I}{2}\left(\left(\frac{d\Theta}{dt}\right)^{2}+\left(\frac{d\Phi}{dt}\right)^{2}+
2\cos\Theta\frac{d\Phi}{dt}\frac{d\Psi}{dt}+\left(\frac{d\Psi}{dt}\right)^{2}
\right).
\]
In mechanics of gyroscopic systems  $\Phi, \Theta, \Psi$ are referred to respectively
as the precession, nutation and rotation angles. This, of course, has nothing to do with
our object, i.e., homogeneously deformable two-dimensional body; such a body has
only one rotational degree of freedom. The analogy is formal,
 nevertheless instructive and effective in the computational sense. The idea has to do
with the "concentric" parametrization of $\mathbb{R}^{4}$. As mentioned, the Cartesian
variables $x$, $y$, $z$, $u$, i.e., matrix elements of the configuration matrix $\phi$,
are non-effective when investigating deformations. This was just the reason to use the two-polar
decomposition and the corresponding coordinates $(D_{1}, D_{2}, \varphi, \psi)$ or  
$(\alpha, \beta, \varphi, \psi)$. The two "radii" $(D_{1}, D_{2})$ or $(\alpha, \beta)$
have to do with the purely scalar deformation; $(\varphi, \psi)$
(equivalently $(\eta, \gamma)$) are angular variables of compact topology (orientation
of deformations in the physical space and in the body). The "concentric" parametrization
consists in encoding the possibility of unbounded motion in the radial variable in 
$\mathbb{R}^{4}$,
\[
\rho=\sqrt{r}=\sqrt{x^{2}+y^{2}+z^{2}+u^{2}}=\sqrt{D_{1}{}^{2}+D_{2}{}^{2}}=
\sqrt{\rm{Tr}(\phi^{T}\phi)}=\sqrt{{\rm Tr} G},
\]
where the symbol $G$ is used for the Green deformation tensor expressed in the Cartesian
coordinates. More geometrically, we are dealing here with the deformation invariant:
\[
\rho=\sqrt{\eta^{AB}G_{AB}}=\sqrt{g_{ij}\phi^{i}{}_{A}\phi^{j}{}_{B}\eta^{AB}},
\]
$g$, $\eta$ denotes respectively the spatial and material (reference) metric tensors.

Degrees of freedom orthogonally transversal to the radial variable $\rho$ 
(or equivalently $r$) describe the geometrically bounded aspect of motion. Those
modes of motion are encoded in the concentric spheres in $\mathbb{R}^{4}$, in 
particular, in the unit sphere given by equation $\rho=1$, i.e., $r=1$. But it
is well-known that the group ${\rm SU}(2)$, i.e., the group of unitary unimodular matrices and
the covering group of ${\rm SO}(3, \mathbb{R})$, may be naturally identified with the
unit sphere $S^{3}(0, 1)\subset\mathbb{R}^{4}$. And in this way this sphere may 
be parametrized with the use of the Euler angles $\Phi$, $\Theta$, $\Psi$. The parametrization of 
$\mathbb{R}^{4}$ with the use of variables $(\rho, \Phi, \Theta, \Psi)$ or
$(r, \Phi, \Theta, \Psi)$ is rather nonusual, however well-suited to the description 
of the three-dimensional rigid body with imposed dilatations or, as we see, to the 
description of the two-dimensional homogeneously deformable body. In other applications
one uses rather spherical systems of coordinates in $\mathbb{R}^{4}$, e.g.,
$r$, $\lambda$, $\mu$, $\nu$,
where
\begin{eqnarray*}
x^{1}&=&r\sin\lambda\cos\mu\cos\nu,\\
x^{2}&=&r\sin\lambda\cos\mu\sin\nu,\\
x^{3}&=&r\sin\lambda\sin\mu,\\
x^{4}&=&r\cos\lambda.
\end{eqnarray*}

Let us mention that the isotropic harmonic oscillator may be described obviously in terms
of those variables, and the expression of Hamiltonian through the action variables
$J_{r}$, $J_{\lambda}$, $J_{\mu}$, $J_{\nu}$, in analogy to (\ref{EQ3.1a}) below, is given by
\begin{equation}\label{EQ2.2c}
H=\omega(2J_{r}+J_{\lambda}+J_{\mu}+J_{\nu}),
\end{equation}
where the degeneracy, i.e., the resonance between $J_{r}$, $J_{\lambda}$, $J_{\mu}$, $J_{\nu}$
is explicitly seen.

One can also use certain mixed type parametrizations in $\mathbb{R}^{4}$, e.g., representing
it as $\mathbb{R}^{3}\times \mathbb{R}$, $\mathbb{R}^{2}\times \mathbb{R}^{2}$ and taking
spherical coordinates in $\mathbb{R}^{3}$ or polar ones in one or two copies of $\mathbb{R}^{2}$.
In all such coordinate systems the isotropic harmonic oscillator is separable and this is
some aspect of its very high, total degeneracy.

However, it is hard to realize a wider class of realistic applications of these
coordinates, e.g., in elastic and similar problems. Unlike this, the apparently exotic
parametrization in terms of the "radial distance" $\rho$ and "Euler angles"
$\Phi, \Theta, \Psi$ offers certain models of potentials which are both separable 
and qualitatively physical.

We have quoted the general St\"{a}ckel-separable Hamiltonian in the variables 
$(r, \vartheta, \varphi, \psi)$ (\ref{EQ2.2f}). It is doubly isotropic when
the shape functions $V_{\eta}$, $V_{\gamma}$ are put as constants. Obviously,
the corresponding terms $V_{\eta} / \cos^{2}(\vartheta / 2)$, 
$V_{\gamma} / \sin^{2}(\vartheta / 2)$ may be simply included into
$V_{\vartheta}(\vartheta)$. We have the following four constants of motion in involution,
responsible for separability:
\begin{itemize}
\item[$\bullet$] $p_{\varphi}, p_{\psi}$, i.e., equivalently $p_{\eta}, p_{\gamma}$,
\item[$\bullet$] $h_{\vartheta}=\frac{1}{2\mu}\frac{1}{\sin^{2}\vartheta}\left(p_{\varphi}{}^{2}
+p_{\psi}{}^{2}+2p_{\varphi}p_{\psi}\cos\vartheta\right)+
\frac{2}{\mu}p_{\vartheta}{}^{2}+V_{\vartheta}(\vartheta)$,
\item[$\bullet$] $H=T+V=H_{r}+\frac{h_{\vartheta}}{r}$,
where, however, the two indicated terms in $H$, namely
\[
H_{r}=\frac{2}{\mu}r p_{r}{}^{2} +V_{r}(r), \quad \frac{h_{\vartheta}}{r}
\]
are not constants of motion when taken separately.
\end{itemize}
The term $V_{r}$ stabilizes the radial mode of motion which without this term
would be unbounded, therefore physically non-applicable in elastic problems.
The term $V_{\vartheta}$ is responsible for the shear dynamics. Let us stress 
that in spite of the "angular" character of $\vartheta$ the shear mode
of motion is also non-compact. It is just seen from the fact that the shear is 
algebraically expressed by the quantity $\tan\vartheta$, which is unbounded.
Therefore, in certain problems some non-constant expression for $V_{\vartheta}$
is also desirable. Even if we use $V_{r}$ proportional to $r=\rho^{2}$,
any model with non-vanishing $V_{\vartheta}$ introduces some anharmonicity.
Particularly interesting is the following simple model:
\begin{equation}\label{EQ2.2h}
V=V_{r}(r)+\frac{V_{\vartheta}(\vartheta)}{r}=\frac{C}{2}r+\frac{2C}{r\cos\vartheta}=
C\left(\frac{1}{D_{1}D_{2}}+\frac{D_{1}{}^{2}+D_{2}{}^{2}}{2}\right).
\end{equation}
The model is perhaps phenomenological and academic, however, from the "elastic"
point of view it has very physical properties: it prevents the collapse to the point or
straight-line, because the term $1 / D_{1}D_{2}$ is singularly repulsive
there, and at the same time it prevents the unlimited expansion, because
the harmonic oscillatory term
$C(D_{1}{}^{2}+D_{2}{}^{2}) / 2=C(\alpha^{2}+\beta^{2}) / 2$
grows infinitely then. There is a stable continuum of relative equilibria
at the non-deformed configurations when $D_{1}=D_{2}=1$. Expansion along some
axis results in contraction along the perpendicular axis, because 
\[
\frac{\partial^{2}V }{\partial D_{1}\partial D_{2}}>0
\]
at $D_{1}=D_{2}=1$. This qualitatively physical potential of nonlinear hyperelastic
vibrations is separable, therefore, at the same time it is also in principle
analytically treatable. Its structure seems to suggest some three-dimensional
models with the attractive harmonic term proportional to 
$(D_{1}{}^{2}+D_{2}{}^{2}+D_{3}{}^{2})$ and some collapse-preventing
term, e.g., one proportional to $(D_{1}D_{2}D_{3})^{-p}$ or $(D_{1}D_{2})^{-p}+
(D_{3}D_{1})^{-p}+(D_{2}D_{3})^{-p}$, $p>0$, however, there is no chance then for 
separability and integrability.

In the chapter below we begin with some problems concerning the harmonic
oscillator,
\begin{eqnarray}\label{EQ3.1}
V(\alpha ,\beta)&=&\frac{C}{2} \left(\alpha ^{2}+\beta ^{2}\right)
=\frac{C}{2}\left(D_{1}{}^{2}+D_{2}{}^{2}\right)\nonumber\\
&=&\frac{C}{2}(x^{2}+y^{2}+z^{2}+u^{2})=\frac{C}{2}{\rm Tr}\left(\phi^{T}\phi\right),
\ C>0.
\end{eqnarray}
and then discuss some natural anharmonic modifications.

\section{Harmonic oscillator and certain anharmonic alternatives}

The expressions $J_{\alpha }$, $J_{\beta }$ depend on potentials
$V_{\alpha }(\alpha )$, $ V_{\beta }(\beta )$, respectively. After
specifying the form of these potentials we can obtain the Hamilton
function $H$ as some function of our action variables, i.e., 
$H=E(J_{\alpha}, J_{\beta}, J_{\varphi}, J_{\psi})$. We
can find the explicit dependence of the energy $E$ on the action
variables and the possible further degeneracy. We will also
perform the usual Bohr-Sommerfeld quantization procedure for our
model.

Hence, we consider the model of the harmonic oscillator potential 
(\ref{EQ3.1}). Some physical comments are necessary here. Namely, the potential 
(\ref{EQ3.1}) describes only the attractive forces which prevent the
unlimited expansion of the body. Its non-physical feature is that
it does not prevent the collapse, i.e., the contraction to the
null-dimensional singularity. It attracts to the configuration
$D_{1}=D_{2}=0$ instead than to the non-deformed state $D_{1}=D_{2}=1$.
Nevertheless, the model may be useful in some range of initial
conditions. Except the subset of measure zero in the manifold of
those conditions, the collapse to $D_{1}D_{2}=0$ is prevented
by the centrifugal repulsion. And the collapse missbehaviour of 
(\ref{EQ3.1}) is not very malicious when the system is discrete.
Obviously, (\ref{EQ2.2a}) and (\ref{EQ3.1}) describe the isotropic
harmonic oscillator in $\mathbb{R}^{4}$ or the quadruple of identical
one-dimensional oscillators in $\mathbb{R}$. In this sense the solution
is obvious and a priori known. Nevertheless, the model is a useful step towards 
investigating more realistic ones. And another point is very important.
Namely, the very strong degeneracy of this model has to do, as usually,
with the separability of the Hamilton-Jacobi equation in several coordinate
systems.

After some calculations we obtain the dependence of  the energy
$E= E_{\alpha }+ E_{\beta }$ on the action variables as follows:
\begin{equation}
E=\frac{\omega}{4 \pi}\left[
4J+|J_{\varphi}-J_{\psi}|+|J_{\varphi}+J_{\psi}|\right], \ J=J_{\alpha }+J_{\beta},
\end{equation}
where $\omega=\sqrt{C / \mu}$ and 
\begin{eqnarray*}
E_{\alpha}&=&\frac{\omega}{4 \pi }\left(4J_{\alpha}+|J_{\varphi }-J_{\psi}|
\right),\\
E_{\beta }&=&\frac{\omega}{4 \pi}\left(4J_{\beta}+|J_{\varphi }+J_{\psi}| \right).
\end{eqnarray*}

Then performing the Bohr-Sommerfeld quantization procedure, i.e.
supposing that $J=nh,\ J_{\varphi }= mh$, $J_{\psi }=lh$, where $h$
is the Planck constant and $n=0, 1, \dots\ $; $ m,l=0,\pm1,\ldots
$, we obtain the energy spectrum in the following form:
\begin{equation}\label{EQ3.2}
E=\frac{1}{2}\hbar\omega\left[ 4n+|m-l|+|m+l|\right].
\end{equation}
We may rewrite this formula as follows:
\begin{itemize}
\item[$(i)$] if $|m|>|l|$, then $m^{2}>l^{2}$ and
\begin{equation}\label{EQ3.21a}
E=\hbar\omega\left(2n\pm m\right),
\end{equation}
\item[$(ii)$] if $|m|<|l|$, then $m^{2}<l^{2}$ and
\begin{equation}\label{EQ3.22a}
E=\hbar\omega\left(2n\pm l\right),
\end{equation} 
\item[$(iii)$] if $|m|=|l|$, then $m^{2}=l^{2}$ and
\begin{equation}\label{EQ3.23a}
E=\hbar\omega\left(2n \pm m\right)=\hbar\omega\left(2n \pm l\right).
\end{equation}
\end{itemize}

And similarly, on the purely classical level of the action variables we have
the following formulas:
\begin{itemize}
\item[$(i)$] in the phase space region where $|J_{\varphi}|>|J_{\psi}|$:
\begin{equation}\label{EQ3.21}
E=\frac{\omega}{2\pi}\left(2J\pm J_{\varphi}\right)=
\frac{\omega}{2\pi}\left(2J_{\alpha}+2J_{\beta}
\pm J_{\varphi}\right),
\end{equation}
\item[$(ii)$] in the region where $|J_{\varphi}|<|J_{\psi}|$:
\begin{equation}\label{EQ3.22}
E=\frac{\omega}{2\pi}\left(2J\pm J_{\psi}\right)=
\frac{\omega}{2\pi}\left(2J_{\alpha}+2J_{\beta}
\pm J_{\psi}\right),
\end{equation}
\item[$(iii)$] on the submanifold  where $J_{\varphi}=J_{\psi}$:
\begin{equation}\label{EQ3.23} 
E=\frac{\omega}{2\pi}\left(2J\pm J_{\varphi}\right)=\frac{\omega}{2\pi}
\left(2J\pm J_{\psi}\right).
\end{equation}
\end{itemize}

The total degeneracy of the doubly invariant model with the potential
(\ref{EQ3.1}) is a priori obvious because in coordinates $(x, y, z, u)$
it is explicitly seen that we deal with four-dimensional isotropic
harmonic oscillator (equivalently--with the quadruple of identical
non-interacting oscillators). If we use coordinates $(D_{1}, D_{2}, \varphi, \psi)$,
or equivalently $(\alpha, \beta, \varphi, \psi)$, then the total degeneracy 
is visualized by the fact that the action variables $J_{\alpha}$, $J_{\beta}$, 
$J_{\varphi}$, $J_{\psi}$ enter (\ref{EQ3.21}) with integer coefficients, $ J_{\psi}$
with the vanishing one. Similarly in (\ref{EQ3.22}) they are also combined 
with integer coefficients, but now the coefficient at $J_{\varphi}$ does
vanish. The third case (\ref{EQ3.23}) is, so-to-speak, the seven-dimensional "separatrice" 
submanifold. The existence of those regions with various expressions for the
functional dependence of energy on the action variables is due to the fact that 
the coordinate system $(D_{1}, D_{2}, \varphi, \psi)$ is not regular in the
global sense and has some very peculiar singularities. Nevertheless, it is just those
coordinates that are more natural and physically lucid in dynamical problems. 

The quasiclassical degeneracy of the Bohr-Sommerfeld energy levels is due
to the fact that the quantum numbers may be combined in a single one, although
in slightly different ways in three possible ranges. Let us observe that in (\ref{EQ3.21a})
the quantum number $l$ still does exist although does not explicitly occur in the
formula for $E$. It runs the range $|l|<|m|$ and labels quasiclassical states within the 
same energy levels. And analogously in the remaining cases (\ref{EQ3.22a}), (\ref{EQ3.23a}).
The action variables $J_{\varphi}$, $J_{\psi}$ and the corresponding quantum numbers $m, l$
take symmetrically the positive and negative values, thus, as a matter of fact, the
ambiguity of signs in the above formulas (\ref{EQ3.21a})--(\ref{EQ3.23a}) does not matter
when the values of energy in stationary states are concerned. Nevertheless, this
ambiguity is essential for classical trajectories, namely, for different signs the orbits
or rather their angular cycles are "swept" in opposite directions.

Let us observe that the formulas (\ref{EQ3.21})--(\ref{EQ3.23}) resemble the action-angle
description of the two-dimensional isotropic harmonic oscillator in terms of usual
polar coordinates $(r, \varphi)$ on $\mathbb{R}^{2}$. Namely, the Cartesian formula
\begin{equation}
E=\omega(J_{x}+J_{y})
\end{equation}
is then alternatively reformulated as
\begin{equation}\label{EQ3.1a}
E=\omega(2J_{r}+J_{\varphi}).
\end{equation}
The ratio $2:1$ of coefficients is due to the fact that the total angular rotation
in the $\varphi$-variable is accompanied by the exactly two total cycles of "libration"
in the $r$-variable. The analogy is neither accidental nor superficial. For the deformative motion
the deformation invariants $D_{1}$, $D_{2}$, i.e., stretchings, are analogues to the radial
variable $r$, whereas the two-polar angles $\varphi$, $\psi$ describing the spatial
and material orientation of stretchings play a role similar to the polar angle $\varphi$
in material point dynamics on $\mathbb{R}^{2}$ (do not confuse--the symbol $\varphi$ is used
in two different meanings). This is just the reason for the $2:1$ ratio in (\ref{EQ2.2c}) and
(\ref{EQ3.21})--(\ref{EQ3.23}). 

Let us now review certain still isotropic, but anharmonic modifications of the harmonic model
of affine vibrations (\ref{EQ2.2a}) and (\ref{EQ3.1}). They are based on the use of variables 
$(\alpha, \beta, \varphi, \psi)$ or $(\rho, \vartheta, \varphi, \psi)$.
The corresponding potentials are given by
\begin{equation}\label{EQa}
V(\alpha, \beta)=\frac{C}{2}\left(\alpha^{2}+\frac{4}{\alpha^{2}}\right)+
\frac{C}{2}\beta^{2}=\frac{C}{2}\left(\alpha^{2}+\beta^{2}\right)+\frac{2C}{\alpha^{2}},
\end{equation}
\begin{equation}\label{EQb}
 V(\rho, \vartheta)=\frac{C}{2}\left(\rho^{2}+\frac{4}{\rho^{2}}\right)+
\frac{2C}{\rho^{2}}\tan^{2}\frac{\vartheta}{2}=\frac{C}{2}\rho^{2}+\frac{2C}{\rho^{2}}
\frac{1}{\cos^{2}\frac{\vartheta}{2}},
\end{equation}
where in both formulas $C$ denoting some positive constant.

Using the former symbols we have
\[
V_{\alpha}=\frac{C}{2}\left(\alpha^{2}+\frac{4}{\alpha^{2}}\right), \ V_{\beta}=
\frac{C}{2}\beta^{2}, \ 
V_{r}=\frac{C}{2}r, \ V_{\vartheta}=
\frac{2C}{\cos^{2}\frac{\vartheta}{2}}.
\]

An important peculiarity of these models is that they have the stable equilibria
in the natural configuration $D_{1}=D_{2}=1$, so they are viable from the elastic
point of view. And both of them are separable ((\ref{EQa}) in the obvious additive sense),
therefore, the corresponding Hamiltonian systems are integrable.

One can explicitly calculate the action variables that correspond to (\ref{EQa}) and (\ref{EQb}), i.e., $(J_{\alpha}, J_{\beta}, J_{\varphi}, J_{\psi})$ and $(J_{r}, J_{\vartheta}, J_{\varphi}, J_{\psi})$. 
They are some functions of the separation constants (one of them is the energy $E$). Eliminating other constants one obtains the expression of $E$, or more precisely, of the Hamiltonian $H$, as a function of action variables.

For the model (\ref{EQa}) one obtains
\[
E=\frac{\omega}{4\pi }\left(4(J_{\alpha}+J_{\beta})+|J_{\varphi}+J_{\psi}
|+\sqrt{64\mu \pi^{2}C+(J_{\varphi}-J_{\psi})^{2}}\right),
\]
where, as usually, we denote
\[
\omega=\sqrt{\frac{C}{\mu}}.
\]
It is seen that the collapse-preventing term $C / \alpha^{2}$ in $V_{\alpha}$
partially removes the degeneracy. Evidently, there is no longer resonance between
$\varphi$ and $\psi$. The resonance between $\alpha$ and $\beta$ obviously survives;
their conjugate actions $J_{\alpha}$, $J_{\beta}$ enter the energy formula through the
rational combination $J=J_{\alpha}+J_{\beta}$ and the corresponding frequencies are equal:
\[
\nu_{\alpha}=\nu_{\beta}=\frac{\omega}{\pi}.
\]
We use here the standard formulas:
\[
\nu_{\alpha}=\frac{\partial E}{\partial J_{\alpha}}, \quad 
\nu_{\beta}=\frac{\partial E}{\partial J_{\beta}}, \quad
\nu_{\varphi}=\frac{\partial E}{\partial J_{\varphi}}, \quad
\nu_{\psi}=\frac{\partial E}{\partial J_{\psi}}.
\]
There are two phase-space regions given respectively by $J_{\varphi}+J_{\psi}>0$
and $J_{\varphi}+J_{\psi}<0$. In any of these regions there is a resonance
between $\gamma=\varphi+\psi$ and $\alpha$, $\beta$. This is seen from the formulas
\[
J_{\varphi}=J_{\eta}+J_{\gamma}, \quad J_{\psi}=-J_{\eta}+J_{\gamma}.
\]
In the mentioned regions we have respectively 
\[
E=\frac{\omega}{4\pi }\left(4J_{\alpha}+4J_{\beta}\pm 2J_{\gamma}
+\sqrt{16\mu \pi^{2}C+J_{\eta}{}^{2} }\right).
\]
This implies the following independent resonances:
\[
\nu_{\alpha}-\nu_{\beta}=0, \quad \nu_{\alpha}\mp 2\nu_{\gamma}=0
\]
or, equivalently,
\[
\nu_{\alpha}-\nu_{\beta}=0, \quad \nu_{\beta}\mp 2\nu_{\gamma}=0.
\]
Therefore, in any of the mentioned regions, where $J_{\gamma}>0$ or $J_{\gamma}<0$,
the system is twice degenerate and the closures of its trajectories are two-dimensional
isotropic tori in the eight-dimensional phase space.

Using the primary variables $\varphi$, $\psi$, we have the following expressions for 
$\nu_{\varphi}$, $\nu_{\psi}$:
\begin{eqnarray*}
\nu_{\varphi}&=&\frac{\omega}{4\pi}\left(\pm 1+
\frac{2(J_{\varphi}-J_{\psi})}{\sqrt{64\mu\pi^{2}C+(J_{\varphi}-J_{\psi})^{2}}}\right),\\
\nu_{\psi}&=&\frac{\omega}{4\pi}\left(\pm 1+
\frac{2(J_{\psi}-J_{\varphi})}{\sqrt{64\mu\pi^{2}C+(J_{\psi}-J_{\varphi})^{2}}}\right),
\end {eqnarray*}
the $\pm$ signs respectively in the regions where $J_{\varphi}+J_{\psi}>0$ or
$J_{\varphi}+J_{\psi}<0$.
Then, taking into account that
\[
\omega=\pi\nu_{\alpha}=\pi \nu_{\beta}=\pi \nu=
\frac{\partial E}{\partial J},
\]
we have the following degeneracy conditions:
\[
\nu_{\alpha}-\nu_{\beta}=0, \quad 
\nu_{\alpha}\mp 2\nu_{\varphi}\mp 2\nu_{\psi}=0,
\]
respectively in the regions where $J_{\alpha}+J_{\beta}>0$ or $J_{\alpha}+J_{\beta}<0$.
Obviously, in the second equation, $\nu_{\alpha}$ may be equivalently replaced by $\nu_{\beta}$.

The corresponding quasiclassical Bohr-Sommerfeld spectrum is given by
\begin{equation}\label{EQ3.6}
E=\frac{1}{2}\hbar \omega\left(4n+|m+l|+\sqrt{(m-l)^{2}+\frac{16C\mu}{\hbar^{2}}}\right).
\end{equation}

Another interesting model is (\ref{EQb}), separable in the variables ($\rho, \vartheta$), i.e.,
equivalently $(r, \vartheta)$. Then we obtain
\begin{eqnarray*}
E&=&\frac{\omega}{4\pi }\left(4(J_{r}+J_{\vartheta})+|J_{\varphi}+J_{\psi}
|+\sqrt{64\mu \pi^{2}C+(J_{\varphi}-J_{\psi})^{2}}\right)\\
&=&\frac{\omega}{4\pi }\left(4(2J_{\rho}+J_{\vartheta})+|J_{\varphi}+J_{\psi}
|+\sqrt{64\mu \pi^{2}C+(J_{\varphi}-J_{\psi})^{2}}\right).
\end{eqnarray*}

Again there is only a two-fold degeneracy and the system is not periodic.
Trajectories are dense in two-dimensional isotropic tori. Degeneracy is described 
by the following pair of independent equations:
\[
\nu_{\rho}-2\nu_{\vartheta}=0, \quad
\nu_{\vartheta}\mp 2\nu_{\varphi}\mp 2\nu_{\psi}=0,
\]
respectively in the phase-space regions where $J_{\varphi}+J_{\psi}>0$ or
$J_{\varphi}+J_{\psi}<0$. Obviously, the second equation may be alternatively replaced by
\[
\nu_{\rho}\mp 4\nu_{\varphi}\mp 4\nu_{\psi}=0.
\]

The corresponding quasiclassical Bohr-Sommerfeld spectrum is given by
\[
E=\frac{1}{2}\hbar \omega\left(4n+|m+l|+\sqrt{(m-l)^{2}+\frac{16C\mu}{\hbar^{2}}}\right),
\]
where the quantum numbers $n$, $m$, $l$, refer respectively to the action variables $J$, 
$J_{\varphi}$, $J_{\psi}$, and the system is twice degenerate. Quasiclassical energy levels are labelled
by two effective quantum numbers, namely, $(4n+m+l)$ and $(m-l)$, and there is also an obvious degeneracy
with respect to the simultaneous change of signs of $m$ and $l$.

Let us mention that some anharmonic potentials independent of $\vartheta$, e.g., the first term in 
(\ref{EQb}), are also of some practical utility as models of a bounded motion. The point is that, as seen
in formula (\ref{EQ2.2x}), the variable $r$ depends both on the area of the body (its "two-dimensional 
volume") and on the shear parameter. Therefore, to be bounded in $r$ implies to be bounded both in 
the "volume" and shear degrees of freedom. Due to the separability, the motion in
$(\varphi, \vartheta, \psi)$-variables is geodetic in the sense of invariant metric tensors on
${\rm SO}(3, \mathbb{R})$ or ${\rm SU}(2)$. And this problem is mathematically isomorphic with the motion of the free 
spherically-symmetric rigid body in the three-dimensional space (purely rotational one, without translations in 
$\mathbb{R}^{3}$).

Another helpful model would be one with $V_{\vartheta}(\vartheta)=A\cos\vartheta$, where $A$ denotes
some constant. The resulting problem is isomorphic with that of the three-dimensional 
heavy top.

It is not excluded that some more general problems from the realm of three-dimensional
gyroscopic dynamics, e.g., symmetric top, might be also of some mathematical usefulness when
studying the two-dimensional affine motion. 

\section{Quantized problems}

Classical dynamical models described above may be easily quantized in the sense of
Schr\"{o}dinger wave mechanics on manifolds. And those rigorously solvable on 
the classical level are so as well on the quantum level.

Let us fix some notation. Let $Q$ be a differential manifold of dimension $n$
with the metric tensor $G$. The components of $G$ with respect to some local coordinates
$q^{1}, \dots, q^{n}$ will be denoted by $G_{ij}$ and the components of the
contravariant inverse of $G$ will be denoted by $G^{ij}$; by definition, 
$G_{ik}G^{kj}=\delta_{i}{}^{j}$. The determinant of the matrix $[G_{ij}]$ will be
briefly denoted by the symbol $|G|$ (no confusion between two its meanings);
it is well-known that, this determinant is an analytic representation of some scalar 
density of weight two; the square root $\sqrt{|G|}$ is a scalar density of weight one.
The invariant measure induced by $G$ will be denoted by $\widetilde{\mu}$; analytically
its element is given by
\[
d\widetilde{\mu}(q)=\sqrt{|G(q)|}dq^{1} \cdots dq^{n}.
\]
Operators of the covariant differentation induced in the Levi-Civita sense by $G$
will be denoted by $\nabla_{i}$. The corresponding Laplace-Beltrami operator $\Delta$ is 
analytically given by
\[
\Delta=G^{ij}\nabla_{i}\nabla_{j}
\]
or explicitly, when acting on scalar fields,
\[
\Delta{\bf\Psi}=\frac{1}{\sqrt{|G|}}\sum_{i,j}\frac{\partial}{\partial q^{i}}
\left(\sqrt{|G|}G^{ij}\frac{\partial {\bf\Psi}}{\partial q^{j}}\right),
\]
${\bf\Psi}$ denoting a twice differentiable complex function on $Q$.

Wave mechanics is formulated in $L^{2}(Q, \widetilde{\mu})$, the space
of square-integrable functions on $Q$ with the scalar product meant as follows:
\[
\langle{\bf\Psi}|{\bf\Phi}\rangle:=\int \overline{{\bf\Psi}}(q){\bf\Phi}(q)d\widetilde{\mu}(q).
\]

The operator $\Delta$ is symmetric with respect to this product, and $\nabla_{i}$
are skew-symmetric. The metric $G$ underlies the classical kinetic energy, therefore, 
the classical energy/Hamiltonian function
\[
H=\frac{\mu}{2}G_{ij}(q)\frac{dq^{i}}{dt}\frac{dq^{j}}{dt}+V(q)=\frac{1}{2\mu}
G^{ij}(q)p_{i}p_{j}+V(q)
\]
becomes the operator
\[
\widehat{H}=-\frac{\hbar}{2\mu}\Delta +V.
\]
Then, denoting and ordering our coordinates $q^{i}$ as $(\varphi, \psi, \alpha, \beta)$
in the Cartesian case we have for explicitly separable isotropic potentials: 

\begin{equation}
[G_{ij}]=\left[\begin{array}{cccc}
\alpha^{2}+\beta^{2} & \beta^{2}-\alpha^{2} & 0 & 0 \\
 \beta^{2}-\alpha^{2} & \alpha^{2}+\beta^{2} & 0 & 0 \\
  0 & 0 & 1 & 0 \\
  0 & 0 & 0 & 1
\end{array}\right],
\end{equation}
\begin{equation}
\widehat{H}=\widehat{H}_{\alpha}+\widehat{H}_{\beta}=-\frac{\hbar
^{2}}{2\mu}\Delta +V(\alpha,\beta),
\end{equation}
where
\begin{equation}
\widehat{H}_{\alpha}=\frac{1}{2\mu}\left(\frac{1}{\alpha^{2}}\left
(\widehat{S}-\widehat{\Sigma}\right)^{2}-\hbar^{2}\left(\frac{\partial^{2}}
{\partial\alpha^{2}}+\frac{1}{\alpha}\frac{\partial}{\partial\alpha}\right)\right)
+V_{\alpha}(\alpha),
\end{equation}
\begin{equation}
\widehat{H}_{\beta}=\frac{1}{2\mu}\left(\frac{1}{\beta^{2}}\left
(\widehat{S}+\widehat{\Sigma}\right)^{2}-\hbar^{2}\left(\frac{\partial^{2}}
{\partial\beta^{2}}+\frac{1}{\beta}\frac{\partial}{\partial\beta}\right)\right)
+V_{\beta}(\beta),
\end{equation}
and $\widehat{S}=(\hbar / i)\partial / \partial\varphi$ is
the spin operator, the generator of spatial rotations about the current
spatial position of the center of mass, whereas
$\widehat{\Sigma}=(\hbar / i)\partial / \partial\psi$ is the
"vorticity" operator, the generator of material rotations.
Operators $\widehat{H}_{\alpha}$, $\widehat{H}_{\beta}$, $\widehat{S}$,
$\widehat{\Sigma}$ are the quantum constants of motion. They also
commute with each other (they represent co-measurable physical
quantities).

Those formulas follow from the expression of $\Delta $ in coordinates 
$(\varphi, \psi, \alpha, \beta)$
\begin{eqnarray}
\Delta {\bf\Psi} &=&\frac{\partial ^{2}{\bf\Psi}}{\partial
\alpha^{2}}+\frac{\partial ^{2}{\bf\Psi}}{\partial
\beta^{2}}+\frac{1}{\alpha} \frac{\partial {\bf\Psi}}{\partial \alpha
}+\frac{1}{\beta} \frac{\partial {\bf\Psi}}{\partial \beta
}+\left(\frac{1}{4\alpha^{2}}+\frac{1}{4\beta^{2}}\right)\left(\frac{\partial
^{2}{\bf\Psi}}{\partial \varphi^{2}}+\frac{\partial ^{2}{\bf\Psi}}{\partial
\psi^{2}}\right)\nonumber \\
&+&\left(\frac{1}{2\beta^{2}}-\frac{1}{2\alpha^{2}}\right)\frac{\partial
^{2}{\bf\Psi} }{\partial \varphi\partial \psi}.
\end{eqnarray}
Separable solutions of the stationary Schr\"{o}dinger equation
$\hat{H}{\bf\Psi} =E\Psi$ have the form:
\begin{equation}
{\bf\Psi} (\varphi, \psi,\alpha
,\beta)=f_{\varphi}(\varphi)f_{\psi}(\psi)f_{\alpha}(\alpha )f_{\beta}(\beta),
\end{equation}
where $f_{\varphi}(\varphi)=e^{im\varphi}$, $f_{\psi}(\psi)=e^{il\psi}$
($m$, $l$ are integers) and $f_{\alpha}(\alpha )$, $\ f_{\beta}(\beta)$ are the
deformative wave functions.

Hence, the stationary Schr\"{o}dinger equation with an arbitrary
potential 
\[
V(\alpha ,\beta )=V_{\alpha }(\alpha )+V_{\beta }(\beta)
\] 
leads after the standard separation procedure to the following
system of one-dimensional eigenequations:
\begin{equation}\label{EQ4.1}
\frac{d ^{2}f_{\alpha}(\alpha)}{d \alpha^{2}}+
\frac{1}{\alpha}\frac{d f_{\alpha}(\alpha)}{d \alpha}
-\frac{\left(m-l\right)^{2}}{4\alpha^{2}}f_{\alpha}(\alpha)+\frac{2\mu}{\hbar^{2}}
\left(E_{\alpha}-V_{\alpha}(\alpha)\right)f_{\alpha}(\alpha)=0,
\end{equation}
\begin{equation}\label{EQ4.2}
\frac{d ^{2}f_{\beta}(\beta)}{d \beta^{2}}+
\frac{1}{\beta}\frac{d f_{\beta}(\beta)}{d \beta}
-\frac{\left(m+l\right)^{2}}{4\beta^{2}}f_{\beta}(\beta)+\frac{2\mu}{\hbar^{2}}
\left(E_{\beta}-V_{\beta}(\beta)\right)f_{\beta}(\beta)=0.
\end{equation}

It is natural to expect that for potentials (\ref{EQ2.1}) the
resulting Schr\"{o}dinger equations should be rigorously solvable
in terms of some standard special functions. The most convenient
way of solving them is to use the Sommerfeld polynomial method
\cite{6,6a}. In this method the solutions are expressed by the usual
or confluent Riemann $P$-functions. They are deeply related to the
hypergeometric functions (respectively usual $F_{1}$ or confluent
$F_{2}$). If the usual convergence demands are imposed, then the
hypergeometric functions become polynomials and our solutions are
expressed by elementary functions. At the same time the energy
levels are expressed by the eigenvalues of the corresponding
operators. There exists some special class of potentials to which
the Sommerfeld polynomial method is applicable. The restriction to
solutions expressible in terms of Riemann $P$-functions is
reasonable, because this class of functions is well investigated
and many special functions used in physics may be expressed by
them. There is also an intimate relationship between these
functions and representations of Lie groups.

Let us now quote some formulas for quantized problems separable in 
coordinates $(r,\Phi, \Theta, \Psi)$ (equivalently 
$(\rho,\Phi, \Theta, \Psi)$), namely, the quantum counterparts
of classical models (\ref{EQ2.2f}). One can easily show that the Laplace
operators take on the form:
\begin{eqnarray*}
\Delta{\bf\Psi}&=&4r\frac{\partial ^{2}{\bf\Psi}}{\partial r^{2}}+8
\frac{\partial{\bf\Psi}}{\partial r}+\frac{1}{r\sin^{2}\vartheta}
\left(\frac{\partial ^{2}{\bf\Psi}}{\partial \varphi^{2}}+
2\cos\vartheta\frac{\partial ^{2}{\bf\Psi}}{\partial\varphi\partial\psi}+
\frac{\partial ^{2}{\bf\Psi}}{\partial \psi^{2}}\right)\\
&+&\frac{4}{r}\left(\frac{\partial ^{2}{\bf\Psi}}{\partial \vartheta^{2}}+
\cot\vartheta\frac{\partial{\bf\Psi}}{\partial \vartheta}\right),
\end{eqnarray*}
i.e.,
\begin{eqnarray*}
\Delta{\bf\Psi}&=&\frac{\partial ^{2}{\bf\Psi}}{\partial \rho^{2}}+\frac{3}{\rho}
\frac{\partial{\bf\Psi}}{\partial \rho}+\frac{4}{\rho^{2}\sin^{2}\Theta}
\left(\frac{\partial ^{2}{\bf\Psi}}{\partial \Phi^{2}}+
2\cot\Theta\frac{\partial ^{2}{\bf\Psi}}{\partial\Phi\partial\Psi}+
\frac{\partial ^{2}{\bf\Psi}}{\partial \Psi^{2}}\right)\\
&+&\frac{4}{\rho^{2}}\left(\frac{\partial^{2}{\bf\Psi}}{\partial \Theta^{2}}+
\cot\Theta\frac{\partial{\bf\Psi}}{\partial \Theta}\right).
\end{eqnarray*}

We assume the doubly isotropic separable potential energy  (\ref{EQ2.2h}), i.e.,
\[
V=V_{r}(r)+\frac{V_{\vartheta}(\vartheta)}{r}=V_{\rho}(\rho)+\frac{V_{\vartheta}
(\vartheta)}{\rho^{2}}.
\]
The corresponding Schr\"{o}dinger equation separates and, taking into account the 
cyclic character of angular variables $\varphi, \psi$, we put
\begin{equation}\label{EQ4.3}
{\bf\Psi}(\varphi, \psi, r, \vartheta)=e^{im\varphi}e^{il\psi}f_{r}(r)
f_{\vartheta}(\vartheta)=e^{im\varphi}e^{il\psi}f_{\rho}(\rho)
f_{\vartheta}(\vartheta),
\end{equation}
where $m, l$ are integers.

Quantum integration constants responsible for this separability are given by operators:
\begin{itemize}
\item  $\widehat{p}_{\varphi}=\frac{\hbar}{i}\frac{\partial }{\partial \varphi}=\widehat{S}$-- spin,
\item  $\widehat{p}_{\psi}=\frac{\hbar}{i}\frac{\partial }{\partial \psi}=\widehat{V}$-- vorticity,
\item  $\widehat{h}_{\vartheta}=\frac{1}{2\mu \sin^{2}\vartheta}\left
(\widehat{p}_{\varphi}{}^{2}+2\cos\vartheta\widehat{p}_{\varphi}\widehat{p}_{\psi}+\widehat{p}_{\psi}{}^{2}
\right)-\frac{4\hbar^{2}}{2\mu}\left(\frac{\partial^{2}}{\partial \vartheta^{2}}+
\cot\vartheta\frac{\partial}{\partial \vartheta}\right)+V_{\vartheta}$,
\item  $\widehat{H}=\widehat{H}_{r}+\widehat{H}_{\vartheta}=\widehat{H}_{r}+\frac{1}{r}\widehat{h}_{\vartheta}=
\widehat{H}_{\rho}+\frac{1}{\rho^{2}}\widehat{h}_{\vartheta}$-- energy,
\end{itemize}
where the "radial energy" is given by
\[
\widehat{H}_{r}=\widehat{H}_{\rho}=-\frac{\hbar^{2}}{2\mu}\left(4r\frac{\partial^{2}}{\partial r^{2}}+
8\frac{\partial}{\partial r}\right)+V_{r}(r)=-\frac{\hbar^{2}}{2\mu}\left(\frac{\partial^{2}}
{\partial \rho^{2}}+\frac{3}{\rho}\frac{\partial}{\partial \rho}\right)+V_{\rho}(\rho).
\]
The four mentioned constants of motion $\widehat{p}_{\varphi}$, $\widehat{p}_{\psi}$, $\widehat{h}_{\vartheta}$,
$\widehat{H}$ are pairwise commuting and therefore they represent co-measurable physical quantities.

\noindent{\bf Warning}: the two indicated contributions to $\widehat{H}$, i.e., $\widehat{H}_{r}$ and $\widehat{H}_{\vartheta}=
\widehat{h}_{\vartheta} / r$ are not constants of motion.
 
The stationary Schr\"{o}dinger equation for the factorized wave function (\ref{EQ4.3}) reduces to
the following pair of ordinary Schr\"{o}dinger equations (Sturm-Lioville equations) for the 
factors depending only on one variable, respectively $\vartheta$ and $r$ (or $\rho$):
\begin{equation}\label{EQ4.4}
\hat h_{\vartheta}f_{\vartheta}=e_{\vartheta}f_{\vartheta},
\end{equation}
\begin{equation}\label{EQ4.5}
\hat{H}_{r}f_{r}+\frac{e_{\vartheta}}{r}f_{r}=Ef_{r}, \quad {\rm i.e}., \quad 
\hat{H}_{\rho}f_{\rho}+\frac{e_{\vartheta}}{\rho^{2}}f_{\rho}=Ef_{\rho}.
\end{equation}

The procedure is first to solve the $\vartheta$-equation and then to 
substitute the resulting eigenvalues $e_{\vartheta}$ to the $r / \rho$-equation.
Then one obtains (at least in principle) the energy levels $E$.

It was mentioned that there exists some strange relationship between the two-polar 
parametrization of ${\rm GL}(2, \mathbb{R})$ and the Euler angles and scale parameters
of rigid body with dilatations. There is some very interesting aspect of this link,
which we noticed first quite accidentally, on the purely analytical level,
before the trivial geometric meaning of this surprise became evident to us.
This artificial detour (wandering about) was due to the fact that by chance we
invented our separating coordinates $(r, \vartheta)$ better $(\rho, \vartheta)$
just where they are rather obscurely hidden, namely as polar parametrization of 
the pair of quantities $(2D_{1}D_{2},  D_{1}{}^{2}-D_{2}{}^{2})$ (\ref{EQ2.2y})--(\ref{EQ2.2z}).

Namely, differential eigenequations (\ref{EQ4.4}), (\ref{EQ4.5}) may be explicitly
written down as follows:
\begin{equation}\label{EQ4.6}
\frac{d^{2}f_{\vartheta}}{d\vartheta^{2}}+\cot\vartheta\frac{d f_{\vartheta}}{d\vartheta}-
\left(\frac{m^{2}+2ml\cos\vartheta+l^{2}}{4\sin^{2}\vartheta}+\frac{\mu}{2\hbar^{2}}(V_{\vartheta}
-e_{\vartheta})\right)f_{\vartheta}=0,
\end{equation}
\begin{equation}\label{EQ4.7}
4r\frac{d^{2}f_{r}}{dr^{2}}+8\frac{d f_{r}}{dr}+
\frac{2\mu}{\hbar^{2}}\left(E-\left(V_{r}+\frac{e_{\vartheta}}{r}\right)\right)f_{r}=0,
\end{equation}
where $m$, $l$ are integers in ${\bf\Psi}$ as coefficients at the angles $\varphi$, $\psi$
in complex exponential functions (eigenfunctions of $\widehat{p}_{\varphi}$, $\widehat{p}_{\psi}$).
Let us now divide by $4$ the nominator and denominator in the bracket expression (\ref{EQ4.6}) 
and formally admit half-integer coefficients.
We can rewrite our equations as follows:
\begin{equation}\label{EQ4.8}
\frac{d^{2}f_{\vartheta}}{d\vartheta^{2}}+\cot\vartheta\frac{d f_{\vartheta}}{d\vartheta}-
\left(\frac{m^{2}+2ml\cos\vartheta+l^{2}}{\sin^{2}\vartheta}+\frac{\mu}{2\hbar^{2}}(V_{\vartheta}
-e_{\vartheta})\right)f_{\vartheta}=0,
\end{equation}
\begin{equation}\label{EQ4.9}
\frac{d^{2}f_{\rho}}{d\rho^{2}}+\frac{3}{\rho}\frac{d f_{\rho}}{d\rho}+
\frac{2\mu}{\hbar^{2}}\left(E-\left(V_{\rho}+\frac{e_{\vartheta}}{\rho^{2}}\right)\right)f_{\rho}=0,
\end{equation}
where now the numbers $m$, $l$ are assumed to run over the set of non-negative integers and
half-integers, i.e., $m,l=0, \frac{1}{2}, 1, \frac{3}{2}, \cdots$.

Let us notice that when there is no purely shear-like potential, i.e., $V_{\vartheta}=0$, then the 
$\vartheta$-equation is just nothing else but the eigenequation for the nutation
$\vartheta$-factor of the stationary states of the spherical top:
\begin{equation}\label{EQ4.10}
\frac{d^{2}f_{\vartheta}}{d\vartheta^{2}}+\cot\vartheta\frac{d f_{\vartheta}}{d\vartheta}-
\left(\frac{m^{2}+2ml\cos\vartheta+l^{2}}{\sin^{2}\vartheta}-\frac{\mu}{2\hbar^{2}}
e_{\vartheta}\right)f_{\vartheta}=0.
\end{equation}
The history of this equation traces back to the Reiche-Rademacher theory of quantum top 
\cite{Re,Ra,19} and to the Wigner theory of irreducible unitary representations of the group
${\rm SU}(2)$ \cite{R, 17,18}, i.e., roughly speaking, to the one-valued and two-valued
irreducible unitary representations of the rotation group $ {\rm SO}(3, \mathbb{R})$. Then
the quantized eigenvalues $e_{\Theta}$ are given by the expression
\[
e_{\Theta j}=\frac{2\hbar^{2}}{\mu}j(j+1)
\]
labelled by non-negative half-integer and integer numbers,
$j=0,1/2, 1, 3/2, \newline \ldots$, i.e., $j\in\{0\}\cup(\mathbb{N}/2)$,
$\mathbb{N}$ denoting the set of naturals.

The corresponding eigenfunctions $d^{j}{}_{ml}(\Theta)$ were found by Wigner
as factors in expressions for the matrix elements of unitary irreducible
representations of ${\rm SU}(2)$,
\[
D^{j}{}_{ml}(\Phi,\Theta,\Psi)=e^{im\Phi}d^{j}{}_{ml}(\Theta)e^{il\Psi}.
\]
Here, as mentioned, $\Phi$, $\Theta$, $\Psi$ denote the Euler angles parametrization of 
${\rm SU}(2)$. Their range is twice larger than the range of Euler angles on the quotient
group ${\rm SO}(3, \mathbb{R})$; this is the reason why the half-integer
quantum numbers do appear.

The celebrated functions $D^{j}{}_{ml}$ appear also as stationary states of the quantized 
spherical free top. Energy levels are then given by
\[
E_{j}=\frac{\hbar^{2}}{2I}j(j+1), \quad j=0, \frac{1}{2}, 1, \frac{3}{2}, \cdots,
\]
$I$ denoting the main moment of inertia, and of course they are $(2j+1)^{2}$-fold degenerate.
The labels of basic $j$-states, $m, l$, are quantum numbers of projections of
the angular momentum respectively on the space-fixed and body-fixed $z$-axes:
\[
\frac{\hbar}{i}\frac{\partial}{\partial \Phi}D^{j}{}_{ml}=\hbar mD^{j}{}_{ml},
\quad \frac{\hbar}{i}\frac{\partial}{\partial \Psi}D^{j}{}_{ml}=\hbar lD^{j}{}_{ml}.
\]
Obviously, $m$, $l$ run over the range $-j, -j+1, \ldots, j-1, j$, jumping by one.
Strictly speaking, in applications concerning the rotational spectra of molecules,
one has to restrict ourselves to integer values of $j$, $m$ and $l$. 
There are however some arguments that perhaps the half integer values might be also
acceptable \cite{1a,3a}.

Let us also mention that $m$, $l$ are good quantum numbers also for a more general
free symmetric top,
not necessarily the spherical one. If $I$, $K$ 
are two main moments of inertia, $I$ doubly degenerate one,
then $D^{j}{}_{ml}$ are still basic eigenfunctions corresponding to the energy levels
\[
E_{j,l}=\frac{\hbar^{2}}{2I}j(j+1)+\hbar^{2}\left(\frac{1}{2I}-\frac{1}{2K}\right)l^{2}.
\]
They are $2(2j+1)$-fold degenerate, namely with respect to the quantum number $m$
and to the sign of $l$.

One can wonder whether such a symmetric free top in three dimensions, 
or more general three-dimensional top with some external potential, first of all one
of the shape $U(\Theta)$ (e.g., heavy top), might be useful as a tool for analyzing
the two-dimensional affinely-rigid body. This is just a question worth to be analyzed.

\section{Quantized harmonic and anharmonic vibrations}

The Schr\"{o}dinger equations from the previous section may be solved only when
the explicit form of potential energy is specified.
It is clear that simple solutions in terms of known special
functions may be expected only when the potential has some
particular geometric interpretation. For example, this is the case
when the corresponding classical problem is degenerate and has
some hidden symmetries.

First let us consider the model of the harmonic oscillator potential
(\ref{EQ3.1}). Applying the Sommerfeld polynomial method we obtain
the energy levels $E=E_{\alpha}+E_{\beta}$ as follows:
\begin{equation}\label{EQ5.1}
E=\frac{1}{2}\hbar\omega\left(4n+4+|m-l|+|m+l|\right),
\end{equation}
where
\begin{equation}
E_{\alpha}=\frac{\hbar\omega}{2}\left (4n_{\alpha}+2+|m-l|
\right), \quad E_{\beta}=\frac{\hbar\omega}{2}\left
(4n_{\beta}+2+|m+l| \right),
\end{equation}
and $\omega=\sqrt{C/\mu}$, $n=n_{\alpha}+n_{\beta}$, $n=0,
1, \dots\ ,$ $m,l=0, \pm1, \dots\ $. We may write:
\begin{itemize}
\item[$(i)$] if $|m|>|l|$, then $m^{2}>l^{2}$ and
\[
E=\hbar\omega\left(2n+2\pm m\right),
\]
\item[$(ii)$] if $|m|<|l|$, then $m^{2}<l^{2}$ and
\[
E=\hbar\omega\left(2n+2\pm l\right),
\]
\item[$(iii)$] if $|m|=|l|$, then $m^{2}=l^{2}$ and
\[
E=\hbar\omega\left(2n+2 \pm m\right)=\hbar\omega\left(2n+2 \pm l\right).
\]
\end{itemize}

After some calculations we obtain the deformative wave functions $f_{\alpha}(\alpha)$
and $f_{\beta}(\beta)$
in the form:
\begin{equation}
f_{\alpha}(\alpha)=\alpha^{\sigma}
\kappa^{\frac{1}{4}+\frac{\sigma}{2}}e^{-\frac{\kappa}{2}\alpha^{2}}F_{2}\left(
-n_{\alpha};1+ \sigma;\kappa\alpha^{2}\right),
\end{equation}
\begin{equation}
f_{\beta}(\beta)=\beta^{\gamma}\kappa^{\frac{1}{4}+\frac{\gamma}{2}}e^{-\frac{\kappa}{2}\beta^{2}}
F_{2}\left(-n_{\beta};1+ \gamma;\kappa\beta^{2}\right),
\end{equation}
where $\sigma =\frac{1}{2}|m-l|$,
$\kappa=\sqrt{C\mu/\hbar^{2}}$, $\gamma
=\frac{1}{2}|m+l|$. 

The constant term $4$ occurying in the rigorous
quantum formula (\ref{EQ5.1}) and absent in the quasiclassical one
(\ref{EQ3.2}) was in principle expected. This resembles the
difference between Schr\"{o}dinger and
Bohr-Sommerfeld-quantized harmonic oscillators. This is an
essentially quantum effect.

In the classical part we mentioned that the harmonic oscillator model,
in spite of its academic character, may have some practical utility, and
besides, it suggests some reasonable anharmonic corrections well suited
to certain of its degeneracy properties. The mentioned corrections reduce
degeneracy in some characteristic way and at the same time the model 
becomes more realistic. On the classical and quasiclassical level we 
discussed the potential (\ref{EQa}), i.e.,
\[
V(\alpha, \beta)=\frac{C}{2}\left(\alpha^{2}+\frac{4}{\alpha^{2}}\right)+
\frac{C}{2}\beta^{2}.
\]
The model may be rigorously solved on the quantum level and one obtains the
following formula for the energy levels:
\begin{equation}\label{EQ5.2}
E=\frac{1}{2}\hbar\omega\left(4n+4+|m+l|+\sqrt{\left(m-
l\right)^{2}+\frac{16C\mu}{\hbar^{2}}}\right).
\end{equation}
The energy in (\ref{EQ5.2}) depends on an integer combination of the
quantum numbers, i.e., $n=n_{\alpha}+n_{\beta}$. The wave functions 
are as follows:
\begin{equation}
f_{\alpha}(\alpha)=\alpha^{\chi}
\kappa^{\frac{1}{4}+\frac{\chi}{2}}e^{-\frac{\kappa}{2}\alpha^{2}}F_{2}\left(
-n_{\alpha};1+ \chi;\kappa\alpha^{2}\right),
\end{equation}
\begin{equation}
f_{\beta}(\beta)=\beta^{\gamma}\kappa^{\frac{1}{4}+\frac{\gamma}{2}}
e^{-\frac{\kappa}{2}\beta^{2}}
F_{2}\left(-n_{\beta};1+ \gamma;\kappa\beta^{2}\right),
\end{equation}
where 
\[
\chi=\frac{1}{2}\sqrt{\left(m-l\right)^{2}+\frac {16C\mu}{\hbar^{2}}}.
\]
It is seen that the formula for the energy levels is structurally "almost"
identical with the quasiclassical one (\ref{EQ3.6}), i.e.,
\[
E=\frac{1}{2}\hbar \omega\left(4n+|m+l|+\sqrt{(m-l)^{2}
+\frac{16C\mu}{\hbar^{2}}}\right).
\]
This is rather typical for systems invariant under "large" symmetry groups
and based on interesting geometric structures. There is a characteristic
shift of energy levels, corresponding to the "null vibrations" of the
harmonic part of the system. Just like on the classical and quasiclassical
levels, the system is twice degenerate and its energy levels are essentially 
controlled by two effective quantum numbers: $n_{\alpha}+n_{\beta}+|m+l|$ 
and $|m-l|$.

Using the formulas (\ref{EQ4.6}), (\ref{EQ4.7}), i.e., (\ref{EQ4.8}), 
(\ref{EQ4.9}), we can also quantize the model (\ref{EQb}), i.e.,
\[
V(r,\vartheta )=\frac{C}{2}\left( r+\frac{4}{r}\right) +\frac{2C}{r}tg^{2}\frac{
\vartheta }{2}.
\]
The expression for the energy levels $E$ is as follows:
\begin{equation}\label{EQ3.2.2}
E=\frac{1}{2}\hbar\omega\left(4n+4+|m+l|+\sqrt{\left(m-
l\right)^{2}+\frac{16C\mu}{\hbar^{2}}}\right),
\end{equation}
where $n=n_{r}+n_{\vartheta}$. The functions $f_{r}(r)$, $f_{\vartheta}(\vartheta)$ 
have the form:
\begin {equation}
f_{r}(r)=r^{-\frac{1}{2}+\varepsilon}\kappa^{\frac{1}{2}+\varepsilon}e^{-\frac{\kappa}{2}r}
F_{2}\left(-n_{r};1+2\varepsilon;\kappa r\right),
\end {equation}
\begin {equation}
f_{\vartheta}(\vartheta)=\left(\cos\frac{\vartheta}{2}\right)^{\chi}\left
(\sin\frac{\vartheta}{2}\right)^{\gamma}F_{1}\left(-n_{\vartheta},1+n_{\vartheta}+
\gamma+\chi;1+\chi;\cos^{2}\frac{\vartheta}{2}\right),
\end {equation}
where
\[
\varepsilon=\frac{1}{2}\sqrt{1+\frac{2\mu}{\hbar^{2}}e_{\vartheta}+\frac{2C\mu}{\hbar^{2}}}, 
\]
\[
e_{\vartheta}=\frac{\hbar^{2}}{8\mu}\left(\left(4n_{\vartheta}+2+
|m+l|+\sqrt{(m-l)^{2}
+\frac{16C\mu}{\hbar^{2}}}\right)^{2}-4-\frac{16C\mu}{\hbar^{2}}\right).
\]

For many physical reasons it would be interesting to discuss the model (\ref{EQ2.2h}),
however, we were not yet successful in solving explicitly the corresponding 
Schr\"{o}dinger equation.

Rigorous solutions for two-dimensional problems may be useful in
microscopic physical problems (vibrations of planar molecules such
as $S_{8}$, $C_{6}H_{6}$) and in macroscopic elasticity (cylinders
with homogeneously-deformable cross-sections). Applications in
dynamics of nanotubes seem to be possible.

The next important thing to be done is a more comprehensive analysis
of the status of analogy with Euler angles and the related complexification
problems. This will be done in a subsequent paper. 
Some introductory analysis is outlined below.

\section{Planar affine body versus spatial rigid body}

It was mentioned above about certain interesting links between mechanics of 
isotropic affine body in two dimension and the dynamics of three-dimensional rigid body,
more precisely, rigid body with imposed dilatations. Only certain analytical aspects, useful in 
calculations, were stressed there. However, the problem is geometrically interesting
in itself and has to do with certain complexification procedures on Lie groups
used as configuration spaces. We shall analyze this problem in more detail in a 
forthcoming paper; here we mention only a few simple analytical relationships.

Let us remind that the metric tensor underlying kinetic energy of the 
planar isotropic affine body was given by
\begin{equation}\label{EQ6.1}
ds^{2}={\rm Tr}\left(d \phi^{T}d \phi \right)=dx^{2}+dy^{2}+dz^{2}+du^{2};
\end{equation}
the corresponding kinetic energy form reads
\begin{equation}\label{EQ6.2}
T=\frac{\mu}{2}{\rm Tr}\left( \frac{d\phi^{T}}{dt}  \frac{d\phi}{dt}\right)=
\frac{\mu}{2}\left( \left( \frac{dx}{dt}\right)^{2}+\left( \frac{dy}{dt}\right)^{2}+
\left( \frac{dz}{dt}\right)^{2}+\left( \frac{du}{dt}\right)^{2}\right),
\end{equation}
where $\mu$ denotes the scalar inertial moment.

For certain reasons it is convenient to use some modified parametrization 
of the two-polar decomposition
\begin{equation}\label{EQ6.3}
\phi=ODR^{-1},
\end{equation}
where $O$, $R$ are proper orthogonal and $D$ is diagonal, namely,
\[
O= \left[\begin{array}{cc}
\cos \frac{\Phi}{2} & -\sin \frac{\Phi}{2} \\
\sin \frac{\Phi}{2} & \cos \frac{\Phi}{2}
\end{array}\right]
, \ D=\left[\begin{array}{cc}
D_{1} & 0 \\
0 & D_{2}
\end{array}\right],
\  R=\left[\begin{array}{cc}
\cos \frac{\Psi}{2} & \sin \frac{\Psi}{2} \\
-\sin \frac{\Psi}{2} & \cos \frac{\Psi}{2}
\end{array}\right]
\]
and $D_{1}={\rm exp}\left( a+b/2\right)$,
$D_{2}={\rm exp}\left(a-b/2\right)$.
It is convenient and instructive from the point of view of our analogies to write
these matrices as:
\[
O={\rm exp}\left( \Phi \frac{1}{2i}\sigma_{2}\right), \
R^{-1}={\rm exp}\left( \Psi \frac{1}{2i}\sigma_{2}\right), \
D={\rm exp}\left( a \frac{1}{2}\sigma_{0}\right){\rm exp}\left( b \frac{1}{2}\sigma_{3}\right),  
\]
where $\sigma_{\nu}$ $(\nu=0, 1, 2, 3)$ are Pauli matrices; more precisely, $\sigma_{{\rm a}}$ $({\rm a}=1, 2, 3)$ 
are "true" Pauli matrices, so
\begin{equation}
\sigma_{0}= \left[\begin{array}{cc}
1 & 0 \\
0 & 1
\end{array}\right]
, \ \sigma_{1}=\left[\begin{array}{cc}
0 & 1 \\
1 & 0
\end{array}\right],
\  \sigma_{2}=\left[\begin{array}{cc}
0 & -i \\
i & 0
\end{array}\right], \
\sigma_{3}= \left[\begin{array}{cc}
1 & 0 \\
0 & -1
\end{array}\right].
\end{equation}

The crucial point for our analogies and links is that the matrices
\begin{equation}
\tau_{{\rm a}}=\frac{1}{2i}\sigma_{{\rm a}}, \quad {\rm a}=1, 2, 3,
\end{equation}
are generators of the group ${\rm SU}(2)$, the universal covering
of ${\rm SO}(3, \mathbb{R})$, with standard commutation rules
\begin{equation}
[\tau_{1}, \tau_{2}]=\tau_{3}, \quad [\tau_{2}, \tau_{3}]=\tau_{1}, \quad [\tau_{3}, \tau_{1}]=\tau_{2}. 
\end{equation}
And similarly, the matrices
\begin{equation}
\widetilde{\tau}_{1}=i\tau_{1}, \quad \widetilde{\tau}_{2}=\tau_{2}, \quad \widetilde{\tau}_{3}=i\tau_{3} 
\end{equation}
are generators of ${\rm SL}(2, \mathbb{R})$ with the standard structure constants,
\begin{equation}
[\widetilde{\tau}_{1}, \widetilde{\tau}_{2}]=\widetilde{\tau}_{3}, \quad [\widetilde{\tau}_{2}, \widetilde{\tau}_{3}]=\widetilde{\tau}_{1}, \quad [\widetilde{\tau}_{3}, \widetilde{\tau}_{1}]=-\widetilde{\tau}_{2}. 
\end{equation}
Obviously, the matrix
\begin{equation}
\tau_{0}=\widetilde{\tau_{0}}=\frac{1}{2}\left[\begin{array}{cc}
1 & 0 \\
0 & 1
\end{array}\right]
\end{equation}
generates real dilatations. So, the matrices $\widetilde{\tau}_{\nu}$ generate the group
${\rm GL}(2, \mathbb{R})$, the configuration space of the planar affine body, and $\tau_{\nu}$
generate $\mathbb{R}^{+}{\rm SU}(2)$, the $2:1$ covering of the configuration space of rigid
body with admitted dilatations ("breathing top"). The rough symbol $\mathbb{R}^{+}{\rm SU}(2)$
denotes the manifold of all matrices obtained as products of special unitary matrices by positive
real numbers, $\mathbb{R}^{+}{\rm SU}(2):=\{\lambda u: \lambda\in \mathbb{R}^{+}, u\in {\rm SU}(2)\}$.

In our models of the doubly isotropic planar affine body, with the metric element (\ref{EQ6.1})
we were used rather to parametrize the plane of deformation invariants $(D_{1}, D_{2})$
by $r=\rho^{2}=(D_{1})^{2}+(D_{2})^{2}$ and the angle $\vartheta$ such that 
$\sin \vartheta=\left(D_{1}{}^{2}-D_{2}{}^{2}\right)/\left(D_{1}{}^{2}+D_{2}{}^{2}\right)$ so that the relationships 
(\ref{EQ2.2y})--(\ref{EQ2.2z}) and those following them are satisfied.
However, in models with affinely-invariant kinetic energies the variables $a, b$ as deformation
invariants are more convenient. As mentioned, one can show that
\begin{eqnarray}\label{EQ6.4}
ds^{2}&=& d\rho^{2}+\frac{1}{4}\rho^{2}\left(d\Theta^{2}+d\Phi^{2}+2\cos\Theta d\Phi d\Psi
+ d\Psi^{2}\right)\nonumber \\
&=&\frac{1}{4r}\left(dr^{2}+r^{2}\left(d\Theta^{2}+d\Phi^{2}+2\cos\Theta d\Phi d\Psi+ d\Psi^{2}\right)\right).
\end{eqnarray}
We easily recognize the term characteristic for the spherical top described in terms of the "Euler angles"
$(\Phi, \Theta, \Psi)$ and the term corresponding to the evolution of the invariant $r$,
a kind of "dilatation" (not in a rigorous sense). Using the more geometric variables $a$, $b$ and the auxiliary,
literally dilatational variable
\begin{equation}
\delta=\sqrt{D_{1}D_{2}}={\rm exp}\left(a/2\right),
\end{equation}
we express (\ref{EQ6.4}) as follows:
\begin{eqnarray}\label{EQ6.4a}
ds^{2}&=&\cosh b \ d\delta^{2} +\delta \sinh b \ d\delta db+\frac{1}{4}\delta^{2}\cosh b \ db^{2} \nonumber \\
&+&\frac{1}{4}\delta^{2}\cosh b \left(d\Phi^{2}+\frac{2}{\cosh b}d\Phi d\Psi+d\Psi^{2}\right).
\end{eqnarray}

This is an ugly non-diagonal form; the reason is that $ds^{2}$ is not affinely-invariant,
but only isotropic. The "Euler angles" term is readable, because, as we saw, $(\cosh b)^{-1}=\cos \Theta$.
There are no essential geometric arguments against modifying (\ref{EQ6.4}) by some extra term proportional to
$d\rho^{2}$.

Let us compare these formulas with those for the spherical three-dimen-\newline sional rigid body with dilatations.
More precisely, we write down the formulas on the group $\mathbb{R}^{+}{\rm SU}(2)$, the universal $(2:1)$
covering group of $\mathbb{R}^{+}{\rm SO}(3, \mathbb{R})$ (roughly speaking, the spinorial breathing-rigid-body).
Again the rough symbol $\mathbb{R}^{+}{\rm SO}(3, \mathbb{R})$ denotes the group of all matrices which are products
of proper rotations (special orthogonal matrices) by positive real numbers, $\mathbb{R}^{+}{\rm SO}(3, \mathbb{R}):=\{\lambda A: \lambda\in \mathbb{R}^{+}, A\in {\rm SO}(3, \mathbb{R})\}$.
Then $\phi\in \mathbb{R}^{+}{\rm SU}(2)$ is "Euler-parametrized" as follows:
\begin{equation}\label{EQ6.5}
\phi={\rm exp}(a\tau_{0}){\rm exp}(\Phi\tau_{2}){\rm exp}(\Theta\tau_{3}){\rm exp}(\Psi\tau_{2}).
\end{equation}

More precisely, historical term "Euler angles" is used when the following convention is used:
\begin{equation}\label{EQ6.6}
\widetilde{\phi} '={\rm exp}(a\tau_{0}){\rm exp}(\Phi\tau_{3}){\rm exp}(\Theta\tau_{1}){\rm exp}(\Psi\tau_{3}),
\end{equation}
or similarly, (more popular in textbooks),
\begin{equation}\label{EQ6.7}
\widetilde{\phi} ''={\rm exp}(a\tau_{0}){\rm exp}(\Phi\tau_{3}){\rm exp}(\Theta\tau_{2}){\rm exp}(\Psi\tau_{3}).
\end{equation}
If (\ref{EQ6.5})--(\ref{EQ6.7}) are identified, then, obviously, $(\Phi, \Theta, \Psi)$
in those formulas denote numerically different functions on ${\rm SU}(2)$. Nevertheless, there is no essential
difference between them. What matters is that the $SU(2)$-matrices are factorized into products of three
elements taken from two orthogonal one-parameter subgroups. This is only the question how those three one-parameter subgroups are called (ordered). The non-historical, apparently exotic convention (\ref{EQ6.5}) is
optimally adapted to our programme of exhibiting some links between planar affine body and spatial rigid body.

Namely, let us take the following metric on $\mathbb{R}^{+}{\rm SU}(2)$, underlying the kinetic energy of the spherical breathing top:
\begin{equation}\label{EQ6.8}
ds^{2}={\rm Tr}\left( d\phi^{\dag}d\phi\right),
\end{equation}
where the "$\dag$ - symbol" denotes Hermitian conjugation of matrices. Denoting again:
\begin{equation}\label{EQ6.9}
\delta={\rm exp}\left(a/2\right), \quad \lambda=\delta^{2}={\rm exp}(a),
\end{equation}
we obtain:
\begin{equation}\label{EQ6.10}
ds^{2}=d\delta^{2}+\frac{1}{4}\delta^{2}\left(d\Theta^{2}+d\Phi^{2}+2\cos\Theta d\Phi d\Psi
+ d\Psi^{2}\right),
\end{equation}
i.e., equivalently,
\begin{equation}\label{EQ6.11}
ds^{2}=\frac{1}{4\lambda}\left(d\lambda^{2}+\lambda^{2}\left(d\Theta^{2}+d\Phi^{2}+2\cos\Theta d\Phi d\Psi
+ d\Psi^{2}\right)\right),
\end{equation}
or,
\begin{equation}\label{EQ6.12}
ds^{2}=\frac{1}{4}e^{a}\left(da^{2}+d\Theta^{2}+d\Phi^{2}+2\cos\Theta d\Phi d\Psi
+ d\Psi^{2}\right).
\end{equation}
Obviously, the $\mathbb{R}^{+}$-factor in $\mathbb{R}^{+}{\rm SU}(2)$ is a normal divisor and from the purely 
geometrical point of view of two-side invariant metrics on $\mathbb{R}^{+}{\rm SU}(2)$,
there are no obstacles against modifying $ds^{2}$ by adding an arbitrary correction term
$ds^{2}{}_{corr}=c \ d\delta^{2}$, $c$ being a constant. This means that (\ref{EQ6.10})--(\ref{EQ6.12}) 
may be replaced by
\begin{equation}\label{EQ6.13}
ds^{2}=(1+c)d\delta^{2}+\frac{1}{4}\delta^{2}\left(d\Theta^{2}+d\Phi^{2}+2\cos\Theta d\Phi d\Psi
+ d\Psi^{2}\right),
\end{equation}
\begin{equation}\label{EQ6.14}
ds^{2}=\frac{1}{4\lambda}\left((1+c)d\lambda^{2}+\lambda^{2}\left(d\Theta^{2}+d\Phi^{2}+2\cos\Theta d\Phi d\Psi
+ d\Psi^{2}\right)\right),
\end{equation}
\begin{equation}\label{EQ6.15}
ds^{2}=\frac{1}{4}e^{a}\left((1+c)da^{2}+d\Theta^{2}+d\Phi^{2}+2\cos\Theta d\Phi d\Psi
+ d\Psi^{2}\right).
\end{equation}

Concerning the extra dilatational term in dynamics of the breathing top, cf, e.g., \cite{5b}. Replacing
the real parameter $a$ in (\ref{EQ6.8}), (\ref{EQ6.9}) by imaginary one $ia$, one obtains instead (\ref{EQ6.15})
the following arc element for the two-side invariant Riemannian metric on the unitary group ${\rm U(2)}$:
\begin{equation}\label{EQ6.16}
ds^{2}=\frac{1}{4}\left((1+c)da^{2}+d\Theta^{2}+d\Phi^{2}+2\cos\Theta d\Phi d\Psi
+ d\Psi^{2}\right).
\end{equation}
For some application or just comparison purposes one can admit in (\ref{EQ6.8}), (\ref{EQ6.9})
the general complex parameter $a$. This results in the doubly-invariant Riemannian metric on
$(\mathbb{C}\  / \left\{ 0\right\}){\rm SU}(2)=\mathbb{R}^{+}{\rm U}(2)$.

This was, so-to-speak, "one side" of injecting geometry and dynamics of the "breathing top"
into those of planar affine body (or conversely). There is also another aspect, namely one based
on affinely-invariant metric tensors on ${\rm GL}^{+}(2, \mathbb{R})$ \cite{14}--\cite{16}.
Such metric tensors, of non-definite signature (${\rm SL}(2, \mathbb{R}),
{\rm GL}^{+}(2, \mathbb{R})$ are non-compact, ${\rm SL}(2, \mathbb{R})$ is semisimple,
and ${\rm GL}^{+}(2, \mathbb{R})$ is the direct product of $\mathbb{R}^{+}{\rm SL}(2, \mathbb{R})$) 
are linear combinations of those given by the arc element
\begin{equation}\label{EQ6.17}
ds^{2}={\rm Tr}\left(\Omega^{2}\right)={\rm Tr}\left(\widehat{\Omega}^{2}\right)
\end{equation}
and the purely dilatational correction term
\begin{equation}\label{EQ6.18}
ds^{2}{}_{corr}={\rm Tr}\left(\Omega\right)^{2}={\rm Tr}\left(\widehat{\Omega}\right)^{2},
\end{equation}
where the Lie-algebraic Cartan one-forms $\Omega$, $\widehat{\Omega}$ on ${\rm GL}(2, \mathbb{R})$
are given by the usual formulas:
\begin{equation}\label{EQ6.19}
\Omega=(d\phi)\phi^{-1}, \quad \widehat{\Omega}=\phi^{-1}d\phi=\phi^{-1}\Omega\phi.
\end{equation}
Of course, (\ref{EQ6.17}) is the main, non-degenerate term of signature $(+++-)$. Killing
tensor on ${\rm GL}(2, \mathbb{R})$ is degenerate; the singular direction is that of the
one-dimensional center $\mathbb{R}^{+}{\rm Id}_{2}$. This Killing case corresponds to the ratio $4:(-2)$
of coefficients at (\ref{EQ6.17}), (\ref{EQ6.18}).

For calculations we need the following parametrization of $\phi\in {\rm GL}^{+}(2, \mathbb{R})$,
analogous to (\ref{EQ6.5})
\begin{eqnarray}\label{EQ6.20}
\phi&=&{\rm exp}(a\widetilde{\tau}_{0}){\rm exp}(\Phi\widetilde{\tau}_{2}){\rm exp}(b\widetilde{\tau}_{3}){\rm exp}(\Psi\widetilde{\tau}_{2})\nonumber\\
&=&\delta {\rm exp}(\Phi\widetilde{\tau}_{2}){\rm exp}(b\widetilde{\tau}_{3}){\rm exp}(\Psi\widetilde{\tau}_{2}),
\end{eqnarray}
where, obviously,
\begin{equation}\label{EQ6.21}
\delta={\rm exp}\left(a/2\right)=\sqrt{\lambda}.
\end{equation}
Combining (\ref{EQ6.17}), (\ref{EQ6.18}) with appropriate coefficients (that at the main term (\ref{EQ6.17})
must be non-vanishing), we finally obtain:
\begin{eqnarray}\label{EQ6.22}
ds^{2}&=&(1+c)d\delta^{2}+\frac{1}{4}\delta^{2}\left(db^{2}-d\Phi^{2}-2\cosh b \ d\Phi d\Psi
- d\Psi^{2}\right) \nonumber \\
&=&\frac{1}{4\lambda}\left((1+c)d\lambda^{2}+\lambda^{2}\left(db^{2}-d\Phi^{2}-2\cosh b \ d\Phi d\Psi
- d\Psi^{2}\right)\right)\nonumber \\
&=&\frac{1}{4}e^{a}\left((1+c)da^{2}+db^{2}-d\Phi^{2}-2\cosh b  \ d\Phi d\Psi
- d\Psi^{2}\right).
\end{eqnarray}

The relationship between these formulas (as matter of fact, one formula written in three alternative forms)
and (\ref{EQ6.4}), (\ref{EQ6.4a}), (\ref{EQ6.10})--(\ref{EQ6.12}) is obvious.
Namely, the last four terms in any form of (\ref{EQ6.22}) become the "minus" terms of the spherical top,
when some complexification procedure is performed, i.e., when we put $b=i\Theta$, $\Theta$ being
real. Then, obviously, the last four terms become the spherical top expression with reversed sign,
\begin{equation}\label{EQ6.23}
-d\Theta^{2}-d\Phi^{2}-2\cos\Theta d\Phi d\Psi- d\Psi^{2},
\end{equation}
and no wonder, because ${\rm SL}(2, \mathbb{R})$ and ${\rm SU}(2)$ are two different (and is a sense,
having opposite properties) real forms of the same complex Lie group ${\rm SL}(2, \mathbb{C})$.
The over-all minus term of the Killing metric on ${\rm SU}(2)$ is due to its compactness. Performing
a similar "imaginarization" of $a$, we obtain just the "minus" expression (\ref{EQ6.16}), the doubly invariant
metric on ${\rm U}(2)$. This also expresses the fact that ${\rm GL}^{+}(2, \mathbb{R})$, ${\rm U}(2)$ are two
different real forms of ${\rm GL}(2, \mathbb{C})$.

\section*{Acknowledgements}

The research presented above was supported by the Ministry of Science and
Higher Education grant No 501 018 32/1992 and Institute of Fundamental
Technological Research PAS internal project 203.

\end{document}